\documentclass[fleqn,usenatbib]{mnras}

\usepackage{newtxtext,newtxmath}

\usepackage[T1]{fontenc}

\DeclareRobustCommand{\VAN}[3]{#2}
\let\VANthebibliography\thebibliography
\def\thebibliography{\DeclareRobustCommand{\VAN}[3]{##3}\VANthebibliography}

\usepackage{graphicx}	
\usepackage{amsmath}	
\let\oldAA\AA
\renewcommand{\AA}{\text{\normalfont\oldAA}}

\title[A look into the halos invisible to eROSITA]{The X-ray invisible Universe. A look into the halos undetected by eROSITA}

\author[P. Popesso et al.]{
P. Popesso,$^{1}$\thanks{E-mail: paola.popesso@eso.org}
A. Biviano,$^{2,3}$
E. Bulbul,$^{4}$
A. Merloni,$^{4}$
J. Comparat,$^{4}$
N. Clerc,
Z. Igo,$^{4}$
A. Liu,$^{4}$
S. Driver,$^{6}$
\newauthor
M. Salvato,$^{4}$
M. Brusa,$^{7}$
Y.E. Bahar,$^{4}$
N. Malavasi,$^{8}$
V. Ghirardini,$^{4}$
A. Robotham,$^{5}$
J. Liske,$^{9}$ 
S. Grandis$^{8}$
\\
$^{1}$European Southern Observatory, Karl Schwarzschildstrasse 2, 85748, Garching bei M\"unchen, Germany\\
$^{2}$INAF-Osservatorio Astronomico di Trieste, via G.B. Tiepolo, 11 - Trieste (Italy) I-34143 \\
$^{3}$IFPU Institute for Fundamental Physics of the Universe, via Beirut, 2 - Trieste (Italy) I-34014 \\
$^{4}$Max Planck Institut f\"ur extraterrestrische Physik (MPE), Postfach 1312, D85741, Garching, Germany\\
$^{5}$IRAP-Roche, 9, avenue du Colonel Roche BP 44346 31028 Toulouse Cedex 4\\
$^{6}$ International Centre for Radio Astronomy Research
(ICRAR), University of Western Australia, Crawley, WA
6009, Australia\\
$^{7}$ Dipartimento di Fisica e Astronomia, Universita' di Bologna, Via Gobetti 93/2, 40129, Bologna, Italy\\
$^{8}$ Universita\"ts-Sternwarte Mu\"nchen / Observatorium Wendelstein
Scheinerstraße 1, D-81679 München, Deutschland\\
$^{9}$Universität Hamburg, Department of Physics, Hamburg Observatory, Gojenbergsweg 112, 21029 Hamburg\\
}

\date{Accepted XXX. Received YYY; in original form ZZZ}

\pubyear{2015}

\begin{document}
\label{firstpage}
\pagerange{\pageref{firstpage}--\pageref{lastpage}}
\maketitle

\begin{abstract}
The paper presents the analysis of optically selected GAMA groups and clusters in the SRG/eROSITA X-ray map of eFEDS (eROSITA Final Equatorial Depth Survey), in the halo mass range $10^{13}-5{\times}10^{14}$ $M_{\odot}$ and at $z < 0.2$. All X-ray detections have a clear GAMA counterpart, but most of the GAMA groups in the halo mass range $10^{13}-10^{14}$ $M_{\odot}$ remain undetected. We compare the X-ray surface brightness profiles of the eROSITA detected groups with the mean stacked profile of the undetected low-mass halos at fixed halo mass. Overall, we find that the undetected groups exhibit less concentrated X-ray surface brightness, dark matter, and galaxy distributions with respect to the X-ray detected halos. The mean gas mass fraction profiles are consistent in the two samples within 1.5$\sigma$, indicating that the gas follows the dark matter profile. The low mass concentration and the magnitude gap indicate that these systems are young. They reside with a higher probability in filaments while X-ray detected groups favor the nodes of the Cosmic Web. Because of the lower central emission, the undetected systems tend to be X-ray under-luminous at fixed halo mass and to lie below the $L_X-M_{halo}$ relation. Interestingly, the X-ray detected systems inhabiting the nodes scatter the less around the relation, while those in filaments tend to lie below it.
We do not observe any strong relationship between the system X-ray appearance and the AGN activity. We cannot exclude the role of the past AGN feedback in affecting the gas distribution over the halo lifetime. However, the data suggests that the observed differences might be related to the halo assembly bias.
\end{abstract}




\section{Introduction}
Understanding how many halos are in our Universe is the most crucial test of our cosmological paradigm. So far this test has been limited to the high mass end of the halo mass function in the galaxy cluster regime, where the exponential shape of the mass function results in the maximal sensitivity to cosmological parameters \citep[e.g.][]{Pillepich2012,Mantz2015, Schellenberger2017, Pacaud2018, Pratt2019}. However, massive clusters sample only the 2\% of the virialized dark matter halo population \citep{2007ApJ...671..153Y}. Addressing this fundamental problem would require extending the analysis to the bulk of the halo population, in the galaxy group regime. However, at this mass scale, our knowledge of the number density of halos and of their properties is still very poor.

There are many ways of selecting halos depending on the properties of their visible components, namely the hot gas and galaxy population. The hot gas emits either in the X-rays, through {\it{Bremstrahlung}} radiation and metal line emission, or in the sub-millimeter through spectral distortion of the cosmic microwave background (CMB) through inverse Compton scattering in the Sunayev-Zeldovich effect. Both types of selections have been largely used mainly to detect the galaxy cluster population. However, the sensitivity and the coverage of previous X-ray and SZ telescopes and surveys prevented so far obtaining a more complete census of the bulk of lower mass halos. Indeed, detecting the hot gas of galaxy groups at temperatures below 1-2 keV always represented a challenge for previous X-ray surveys. As a result, the number of known X-ray and SZ galaxy groups is still very low, although low mass halos should by far outnumber the massive clusters \citep{Ponman1996, Mulchaey2000, Osmond2004, Sun2009, Lovisari2015}. 

In addition, being only the tip of the iceberg, the observed galaxy group samples very likely provide only a biased view of the real underlying halo population in terms of global X-ray and SZ properties \citep[see e.g.][]{2021Univ....7..254L,2021Univ....7..139L,2021Univ....7..208G}. For instance, the current paradigm of galaxy formation and evolution postulates that the feedback generated by the super-massive back hole hosted by the central galaxy is capable of releasing a sufficient amount of energy or heat to affect the properties of the gas \citep[see for an extensive review][]{2021Univ....7..142E}. While in galaxy clusters this effect invests only the core region, in low-mass halos it extends to the virial scale. Different implementations of such  feedback might lead to strikingly divergent predictions, varying from exceedingly hot gas-rich groups to systems completely devoid of gas, with a direct effect on the X-ray appearance of these systems on a global scale. In such a scenario, the X-ray selection would, in principle, capture only halos where the effect of the feedback has been marginal \citep{LeBrun2014}. 

Alternatively, the X-ray appearance of a system might be biased by the clustering properties of the underlying halo population. The spatial distribution of dark matter halos primarily depends upon the halo mass \citep[e.g.,][]{Kaiser1984, Efstathiou1988, MoWhite1996}. However, there have been several studies that show that the clustering of halos at fixed halo mass can further depend upon secondary properties related to their assembly history \citep[e.g.,][]{gao_etal05, wechsler06, gao_white07, faltenbacher_white10, dalal_etal08}. This dependence of the halo clustering amplitude on secondary properties other than the halo mass is commonly referred to as \textit{halo assembly bias}, and can be traced back to the fact that halos of the same mass in different environments have different assembly histories and cluster differently. In numerical studies, the assembly bias trend suggests that at large halo masses highly concentrated or old halos cluster weakly as compared to less concentrated or younger halos of the same mass. On the other hand, at low masses the trend inverts, with old halos clustering more strongly than younger ones. The mass at which the inversion takes place is defined as $m*$, which is the scale where the mass fraction in halos has a maximum. Such a threshold in most recent papers appears to be in the low mass halos regime around $10^{13}$ $M_{\odot}$ \citep[see for  instance][]{Ramakrishnan2019, Paranjape2018}. The trend for massive halos is, in fact, qualitatively predicted by simple models of structure formation based on peaks theory \citet{dalal_etal08}, ellipsoidal dynamics \citet{Desjacques2008}, or the excursion set formalism \citet{Musso2012,Castorina2013}. These correlations naturally produce halos with large inner density (or high concentration) that form early and live in more underdense environments as compared to halos of the same mass but with lower inner density, that form late and live in denser environments. In this respect, in the low luminosity and low mass regimes of the galaxy groups, the X-ray selection might favor the detection of highly concentrated halos, with a clear bias towards old halos in underdense or overdense environments, depending on the halo mass threshold $m*$.

A different means of selecting massive halos, such as groups and clusters, is based on the spatial distribution of galaxies and their clustering properties in complete spectroscopic surveys. Differently from the X-ray selection, such a technique, mostly used at the optical wavelength in the local Universe, is very efficient in identifying low mass groups, independently from their hot gas content. In this respect, large and complete galaxy spectroscopic surveys, such as SDSS \citep{blanton2017} and GAMA \citep{Driver2022}, provide the most complete census of optically selected groups and clusters, down to halo masses of $10^{12}$ $M_{\odot}$ at low and intermediate redshift \citep{Lim2017, Tempel2017, Robotham2011, 2007ApJ...671..153Y}. 

The combination of the optically selected samples of groups and clusters and large X-ray surveys proved to be very efficient in determining the average X-ray properties of the underlying halo population down to group mass scale \citep{Anderson2015, rozo2009, rykoff2008}. The stacking in the X-ray data at the position of the optically detected groups allowed to measure the average X-ray scaling relations down to the low mass regime. However, such analysis was possible mainly in the shallow and low spatial resolution ROSAT All Sky Survey (RASS) data, which enabled only an estimate of the mean system X-ray luminosity, with no information on the spatial distribution of the gas and its temperature. More recently the availability of deeper surveys, such as XMM-XXL allowed us to obtain a higher spatial resolution, but the small surveyed area significantly limits the statistics \citep{Crossett2022,Giles2022}. 

The situation will change with the availability of the eROSITA All Sky Survey (eRASS) data. With its $10^5$ galaxy groups, and an immense surveyed area, eRASS will allow overcoming the current limits in terms of statistics, depth and spatial resolution \citep{Predehl2021}. Indeed, the combination of the SDSS and GAMA optically selected groups with the eROSITA detections and stacking analysis will enable to determine which halos the X-ray selection is able to capture and which percentage of the halo population at fixed halo mass remains invisible to eROSITA. In this paper we present a first analysis of the combination of the GAMA and the eROSITA data over the eROSITA Final Equatorial Depth Survey (eFEDS) area \citep{Brunner2022} as a test case for the future analysis on the eRASS data.  

The paper is structured as follows. In Section 2 we provide a description of the GAMA optically selected group sample and of the eFEDS data. In Section 3 we describe the X-ray stacking analysis and its results. In section 4 we present the dynamical analysis of the sample. In Section 5 we analyze several properties of the X-ray-detected and undetected halo samples to identify differences and similarities. In Section 6 we analyze the location of the mean X-ray luminosity per halo mass bin, as obtained from the stacking analysis, in the $L_X-M_{halo}$ relation. Section 7 provides our discussion and interpretation of the results and our conclusions. 

Throughout the paper we assume a Flat $\Lambda$CDM cosmology with $H_0 = 67.74$ km~s$^{-1}$~Mpc$^{-1}$, $\Omega{_m}(z = 0) = 0.3089$ \citep{Planck2016}.

\begin{figure*}
\includegraphics[width=\textwidth]{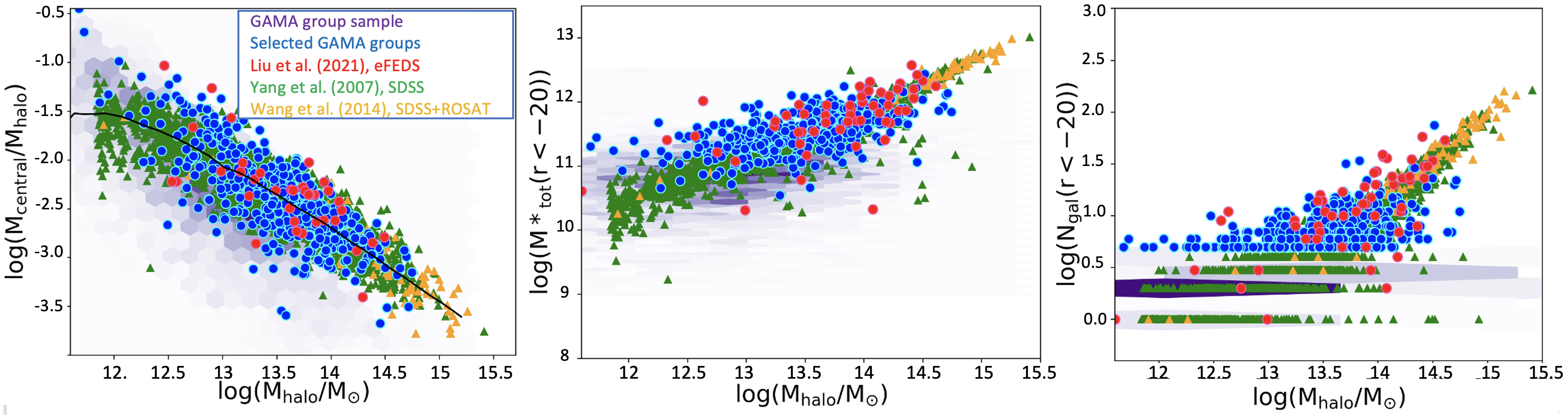}
\caption{Optical properties of the GAMA undetected and the eFEDS detected systems. We compare the two samples with the SDSS group sample of \citet{2007ApJ...671..153Y} in a similar redshift range and with the sample of \citet{Wang2014}, who provide the ROSAT detections of the optically selected groups. {\it{Left panel:}} ratio of the central galaxy stellar mass over the halo mass versus the halo mass. We use $M_{200}$ for all samples, by correcting for the different cosmology when needed. The same halo mass is used in all panels. The purple-shaded region indicates the density distribution of all GAMA groups in the diagram. The blue points indicate the GAMA undetected groups considered in this work. The red points indicate the eFEDS-detected systems. The green triangles show the \citet{2007ApJ...671..153Y} systems while the orange triangles indicate the subsample with a $SNR>5$ ROSAT detections in \citet{Wang2014}. The same color coding is applied to all panels. {\it{Central panel:}} Total stellar mass of the groups versus halo mass based on the GAMA velocity dispersion. The total stellar mass is estimated by considering all system members brighter than $r_{mag}=-20$. {\it{Right panel:}} System richness versus halo mass. The richness is estimated as the number of system members brighter than $r_{mag}=-20$. }
\label{scaling}
\end{figure*}

\section{The data set}
\label{dataset}
\subsection{eROSITA eFEDS data}

We used for this work the public Early Data Release (EDR) eROSITA event file of the eFEDS field \citep{Brunner2022}. The field is observed with an exposure ($\sim$2.5 ks unvignetted) slightly higher than the future eRASS after completion ($\sim$1.6 ks unvignetted). It contains about 11 million events (X-ray photons), detected by eROSITA, over the 140 deg$^2$ area of the eFEDS Performance Verification survey. Each photon is assigned an exposure time using the vignetting corrected exposure map. Photons close to detected sources in the source catalog are flagged. These sources are cataloged as point-like or extended based on their X-ray morphology \citep{Brunner2022}, and they are further classified (e.g. galactic, active galactic nuclei, individual galaxies at redshift z < 0.05, galaxy groups and clusters) using multi-wavelength information \citep{Salvato2022, Vulic2022, LiuTeng2022, LiuAng2022, Bulbul2022}. 

\subsection{The eROSITA X-ray selected group and cluster sample}

We used in this analysis the eROSITA X-ray selected group and cluster sample of \citet{LiuAng2022}. This comprises more than 500 extended objects up to z$\sim$1. As explained in Section 2.4, this is not strictly a flux selected sample. However, according to \citet{LiuAng2022} the catalog reaches a completeness of 40\% down to a flux limit of $1.5 \times 10^{-14}$ erg/s/cm$^2$. Each source is assigned a redshift according to the analysis reported in \cite{Klein2022}, an estimate of the X-ray luminosity within different apertures (300 and 500 kpc and within $R_{500}$\footnote{$R_{\Delta}$ is the radius of a sphere centered on the group center, enclosing a mean density $\Delta$ times the critical density of the Universe at the group redshift.}), and the surface brightness distribution within several radii. An estimate of $M_{500}$\footnote{$G \, M_{\Delta}=\Delta/2 \, H_z^2 \, R_{\Delta}^3$, where $G$ is the gravitational constant and $H_z$ the Hubble constant at the group redshift $z$.} is provided on the basis of the $L_X-M_{500}$ scaling relation of \citet{Lovisari2015}. 

All eFEDS extended sources in the local Universe ($z < 0.2$, which is the redshift window of interest for the presented  analysis) have an optical counterpart in the GAMA sample. Only in a few cases (10\% of the sample), the
association is not clear, because several lower mass groups are at the same redshift and in the region of the X-ray detection. In these cases, we assign to the X-ray system the most massive GAMA group as an optical counterpart. However, due to the uncertainty, the multiple associations are excluded from the analysis. In one case, the GAMA optical counterpart is clear but at a different redshift than the one reported in the eFEDS catalog of \citet{LiuAng2022}, which is above $z=0.2$.

\subsection{The GAMA optically selected group and cluster sample}
The optically selected GAMA group sample \citep{Robotham2011} comprises about 7500 galaxy groups and pairs identified in the spectroscopic sample of the GAMA spectroscopic survey \citep{Driver2022}. The GAMA spectroscopic survey reaches a completeness of $\sim 95\%$ down to the magnitude limit of $r=19.8$. The galaxy groups and pairs are identified through the FoF algorithm described in \cite{Robotham2011}. The detection algorithm has been largely tested on mock catalogs to optimize completeness and contamination levels and to test the ability in reproducing the underlying predicted halo population in the group mass regime. 

Once the galaxy membership is identified, the mean group coordinates and redshift are estimated iteratively for each system. The total mass of the systems ($M_{\mathrm{fof}}$) is estimated from the group velocity dispersion ($\sigma_v$) within a given variable radius \citep[see][for more details]{Robotham2011}. Since this mass is not related to a definite over-density, we derived estimates of $M_{200}$ from the group $\sigma_v$ through the scaling relation of \citet[][Table 1, "AGN gal"]{Munari2013}, 
that was based on cosmological numerical simulations. 
In the following analysis, whenever we refer to the halo mass, unless otherwise specified, we refer to the measure of $M_{200}$. We also derived a measure of $M_{500}$, obtained from the $\sigma_v$-based estimates of $M_{200}$,  using the NFW mass distribution model and the concentration-mass relation of \citet{DM14}

In addition, we cross-matched the catalog of galaxy members with the GAMA catalog of \citet{Bellstedt2021} and \citet{Tailor2011} to retrieve stellar masses and rest-frame photometry of the individual galaxies, respectively.

\subsection{The detected and undetected system samples.}
To understand how many of the halos, from the mass scale of groups to clusters, the X-ray selection at the eFEDS depth is able to capture and how many remain undetected, we apply the following approach:

\begin{itemize}
    \item we define a clean, reliable, and complete optically-selected prior group and cluster sample, able to reproduce the predicted underlying halo mass distribution in the considered halo mass range. We will call this sample the "parent sample" hereafter;
    \item all X-ray detections are removed from the parent sample in order to separate what eROSITA detects from what remains undetected.
\end{itemize}

With this approach, we distinguish between detected and undetected systems. We note that the GAMA galaxy sample only covers a fraction of the eFEDS area, 60 deg$^2$ out of 140 deg$^2$. Necessarily, we limit our X-ray analysis to this 60 deg$^2$ area.

To create the parent sample, we need first to assure the purity and completeness of the GAMA optically selected group and cluster sample. This captures groups and galaxy pairs down to very low halo masses. However, since the halo mass is based on the velocity dispersion, the lower the number of members, the higher the uncertainty of the derived halo mass. Since GAMA is a magnitude-limited survey, such uncertainty is the highest for very poor groups and pairs and it increases as a function of redshift. Thus, to ensure a reliable estimate of the velocity dispersion and derived halo mass, we limit the analysis to the redshift range $0 < z < 0.2$. Such limit enables to have a complete galaxy sample down to $r_{mag} < -20$, which in turn allows estimating also other halo mass proxies, such as the richness, the total stellar mass and the total $r$-band luminosity of the system members without uncertainties due to incompleteness \citep[see also][]{Popesso2005,2007ApJ...671..153Y}. In addition, out of the original GAMA group and cluster sample, we select systems with at least 5 members to ensure a more reliable estimate of the velocity dispersion. 

The resulting group sample follows the known scaling relations between halo mass, as estimated by the velocity dispersion, and other mass proxies (see Fig. \ref{scaling}), also when compared to other group and cluster samples as the one of \citet{2007ApJ...671..153Y} and \citet{Tempel2017} based on the SDSS data in the same redshift range. We point out that there is no segregation in terms of the halo mass of the detected groups with respect to the undetected ones in any of the scaling relation plots. This suggests that the undetected group masses are not biased compared to the detected systems. So the non-detection in the eROSITA data cannot be solely attributed to an overestimate of the undetected group halo masses. Since the richness is a proxy of the halo mass, the applied richness cut implies a cut in halo mass at $\sim 10^{13}$ $M_{\odot}$, as we show below.

\begin{figure}
\includegraphics[width=0.5\textwidth]{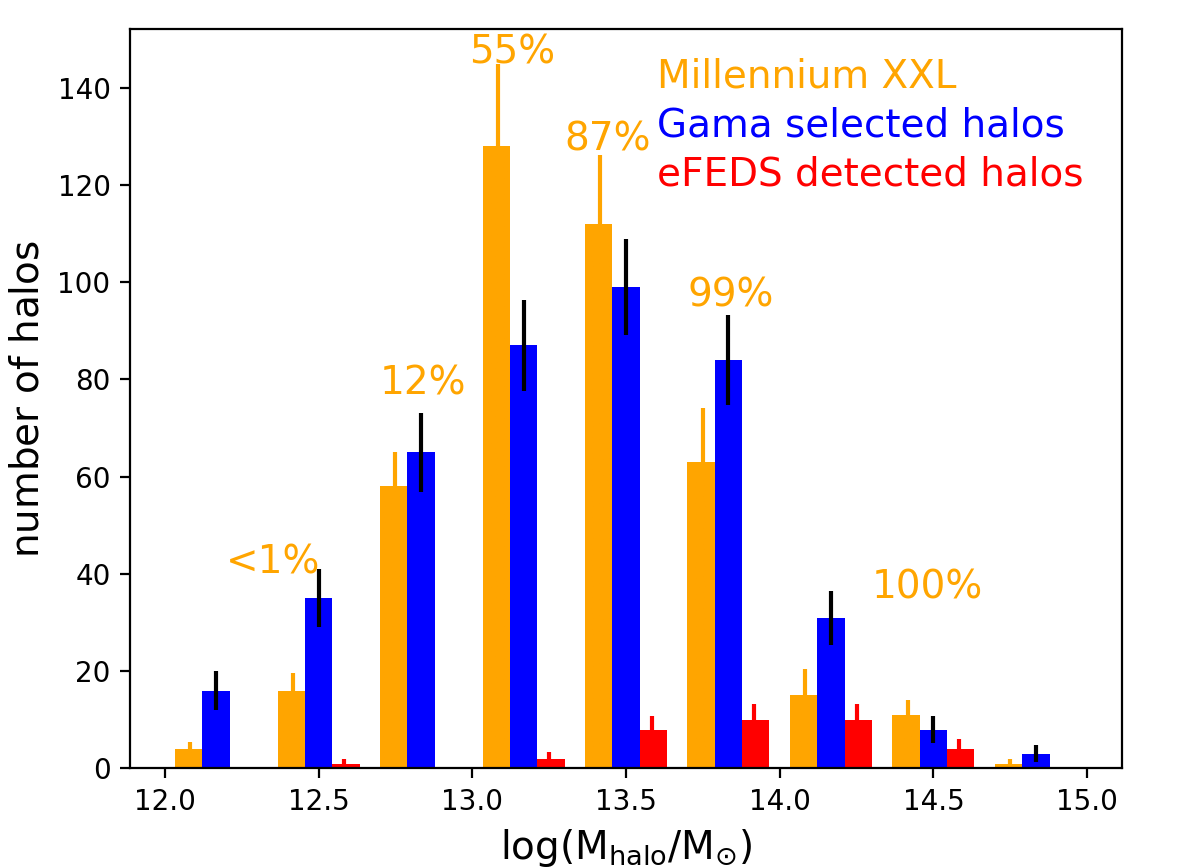}
\caption{Halo mass distribution in the Millennium XXL sample, averaged over ten volumes as large as the one sampled by the overlap of the eFEDS and GAMA area in the redshift range $0 < z < 0.2$ (orange histogram). The red histogram indicates the X-ray detected eFEDS sample in the overlap region of the GAMA and eFEDS area. The blue histogram shows the optically selected GAMA group sample in the same area. The cut in richness $\geq 5$ of galaxy members is applied to all samples for consistency. We also indicate the percentage of halos obtained after the cut in richness with respect to the parent sample in the Millennium XXL sample. Orange, blue, and red histograms are in the same halo mass bins but are displaced along the x-axis for the sake of clarity.}
\label{halomass}
\end{figure}

To verify to what extent and down to which halo mass the considered parent sample is representative of the underlying halo mass distribution, we provide a comparison with the halo mass distribution predicted by the Millennium XXL simulation, which samples a volume large enough to be comparable with the eFEDS observation. To this aim, we use the light cones generated as described in \citet{Smith2017} and available at the Virgo database ( http://virgodb.dur.ac.uk:8080/MyMillennium/). The light cones are provided in two catalogs: one listing the halo mass, radius, coordinates, and observed redshift of the halo population, and one providing coordinates, observed redshift and broadband photometry in few optical bands, including the SDSS $r$-band, of the galaxy population associated to each halo. We select 10 random regions with an area as large as the overlap of the GAMA and eFEDS regions (60 deg$^2$) and in the redshift range $0 < z < 0.2$. For each system, we measure the richness down to the same absolute magnitude limit as in the observed sample, and we apply the same cut at a richness of 5. We average the number counts per mass bin and use the dispersion as a measure of the uncertainty due to cosmic variance. Fig. \ref{halomass} shows the average halo mass distribution of Millennium XXL light cones (orange histogram) and indicates the percentage of halos selected after the cut in richness with respect to the parent population. The cut in richness seriously affects the counts at halo masses of $\sim 10^{13}$ $M_{\odot}$, by selecting only $\sim55\%$ of the halos in the corresponding mass bin. 
Below such limit, the richness cut gets rid of the majority of the low-mass halos, as expected. The halo mass distribution of our parent sample is shown by the blue histogram, which reproduces the Millennium XXL halo mass distribution, within $1.5 \sigma$ in the worst case, down to $10^{13}$ $M_{\odot}$. 
We conclude that the parent sample drawn in this way from the GAMA group sample provides a good representation of the underlying halo population down to halo masses of $10^{13}$ $M_{\odot}$. 

As mentioned earlier, the eROSITA X-ray detection algorithm for extended sources does not lead to a pure flux limit selection \citep[see Section 3.4 for more details and also][]{Clerc2018,Seppi2022}. However, \citet{LiuAng2022} report that a flux limit of $1.5\times10^{-14}$ erg/s/cm$^{-2}$, captures 40\% of the extended sources with $L_{500} > 10^{42}$ erg/s at $z < 0.2$ and with with $L_{500} > 10^{41.5}$ erg/s at $z < 0.1$ \citep[see Fig. 5 of][]{LiuAng2022}. According to \citet{Klein2022}, these correspond to a halo mass threshold of $M_{500}$ of $5\times 10^{13}$ $M_{\odot}$ at $z < 0.2$ and $3\times 10^{13}$ $M_{\odot}$ at $z < 0.1$. The thresholds are roughly consistent with the predictions of \citet{Pillepich2012} (right panel of their Fig. 3). However, one must take into account that, by construction, a flux-limited selection would capture only the upper $L_X$ envelope of the $L_X-M_{500}$ relation at fixed halo mass. Currently, the scatter of the $L_X-M_{500}$ relation at such low halo masses is still unknown because of the low statistics \citep[e.g.][]{2021Univ....7..139L}. This makes it very difficult to estimate the completeness of the X-ray selection in terms of halo mass. 
However, as shown by the red histogram in Fig. \ref{halomass}, the comparison with the halo mass distribution of the local Millennium XXL groups suggests that the eROSITA X-ray selection is able to capture only a very low fraction of the underlying halo population at $5\times 10^{13}$ $M_{\odot}$ at $z < 0.2$ and it does not detect also a fraction of poor clusters with masses at $10^{14}$ $M_{\odot}$. This would indicate that eROSITA extended source catalog in the regime of the galaxy groups, below $L_X \sim 10^{43}$ erg/s and $M_{500} \sim 10^{14}$ $M_{\odot}$, as defined in \citet{2021Univ....7..139L}, captures only a sub-population of the galaxy group family. 

Thus, the questions are: how representative of the underlying group population the X-ray detected sub-sample is? what are the average X-ray properties of the underlying group population? and what physical process might affect the gas properties and the X-ray appearance of groups at fixed halo mass to place then above or below the detection threshold of the eROSITA surveys? To answer these questions we need to stack the undetected systems in the eROSITA map, as a function of the halo mass, to study their mean X-ray properties. Indeed, since all the eFEDS detections have an optical counterpart in the GAMA catalog with the exclusion of a very low percentage of unclear cases (see Section 2.2 for more details), combining the detections and the stacking of all remaining undetected GAMA groups at fixed halo mass provides a census and a clear overview of the mean X-ray properties of local galaxy group population.

For the stacking analysis, in addition to the detected systems, we exclude from the prior sample all optically selected systems closer to an X-ray detection by less than 2.5 Mpc, which is the region we use to study the mean X-ray surface brightness distribution. The additional percentage of systems removed in this way is less than 4\%. In the following analysis, we refer to this sample as the "undetected system sample". Instead, we will call thereafter the eFEDS detected groups and clusters, selected with the same criteria as the optical ones (richness $> 5$), as "detected systems".

Particular attention must be given also to the potential contamination due to point source detection. In this respect, we choose the following approach. Systems with a detected point source falling at a distance between 0.5 to 2 Mpc from the system center are kept and the region (all events) associated to the point source is masked out and not considered in the stacking analysis. When estimating the surface brightness we carefully exclude also the area associated to the detected point source. Groups with an X-ray point source falling within the central 0.5 Mpc from the center are discarded. We identified 45 systems of this category, of which 23 are above the considered mass threshold of $10^{13}$ $M_{\odot}$. In all cases but one, the AGN in the point source catalog is robustly associated on the basis of the X-ray and optical properties to one of the galaxy members. These systems, indeed, have not been identified as "cluster in disguise" by \citet{Bulbul2022} according to the method described in \citet{Klein2022}. The systems of \citet{Bulbul2022} are highly concentrated and likely affected by the contribution of the X-ray emission of the central galaxy AGN. They are miss-classified as point-sources by the eROSITA detection algorithm, although the association to the host galaxy is highly uncertain. 
In our case, the secure association of the galaxy host would suggest that the X-ray signal is mostly due to the nuclear activity of the galaxy member hosting the AGN. Since this would highly contaminate the stacked signal, we exclude these systems from the stacking analysis, but we will include them in the analysis of the global sample properties.
We checked that the halo mass (or halo mass proxy) distribution of the undetected sample does not change significantly after the exclusion of such objects (according to a KS test).

\begin{figure}
\includegraphics[width=0.5\textwidth]{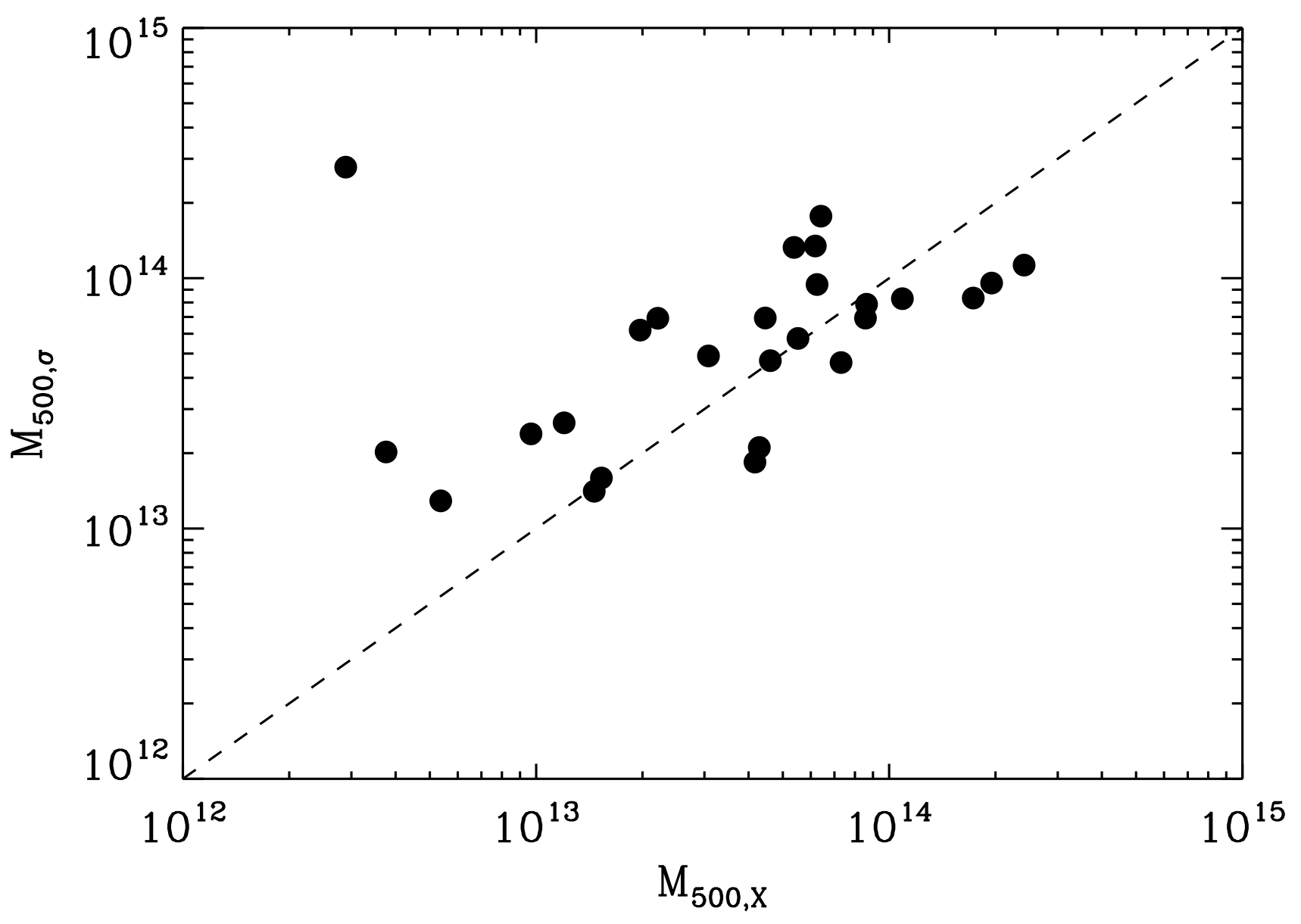}
\caption{$\sigma_v$-based $M_{500}$ estimates vs. X-ray data based $M_{500}$ estimates for 26 of the 32 detected clusters in our sample, for which both estimates are available. The identify line is shown (dashed line).}
\label{f:m500xsv}
\end{figure}

The selection procedure described above leads to a sample of 189 systems at $z<0.2$, with more than 5 spectroscopic members, and $M_{200} \geq 10^{13} M_{\odot}$, of which 32 detected in X-ray and 157 undetected. In Fig.~\ref{f:m500xsv} we show the comparison between the $M_{500}$ estimates obtained from the X-ray data analysis \citet[][see for details]{LiuAng2022}, $M_{500,X}$, that are available for 26 out of the 32 detected systems, and the corresponding estimates obtained from the system velocity dispersion, $M_{500,\sigma}$. 
Apart from an obvious outlier (mentioned in Section 2.2 amd due to a redshift discrepancy), the two mass estimates agree very well, with a median ratio $M_{500,X}/M_{500,\sigma}=0.96$. The outlier in the relation is due to a different redshift assignment of the redMapper algorithm applied by \citep{Klein2022} and the optical group counterpart in the GAMA group sample. 

\section{X-ray stacking analysis}

\subsection{Method}

For the stacking procedure, we adapt the method of \citet{Comparat2022} to our case. We retrieve all the events within the angle subtended by 3 Mpc at the group prior coordinates and redshift. We construct a ‘cube’ of event surrounding each group. So, all such eROSITA events are characterized by the following vector: its position on the sky (R.A., Dec.), the corresponding group redshift ($z_G$), the exposure time ($t_{obs}$), the on-axis telescope effective area as a function of energy (ARF) at the observed energy, the projected distance in physical (kpc, $R_{\mathrm{kpc}}$) and observed (arcsec, $R_{\mathrm{arcsec}}$) units from the group center, and a flag ($f_{ps}$) indicating whether the event is associated to a point source and needs to be masked. To go from number counts to flux, we use the conversion factor $1.624\times 10^{-12}$ erg/cm$^{-2}$ \citep[eg][]{Brunner2022,LiuAng2022}. We, then, select all the events with fluxes corresponding to the rest-frame $0.5-2$ keV at the median redshift of the prior sample, which in all cases is $\sim 0.1$ given the redshift distribution of the prior sample. The exposure times are obtained from the vignetted exposure map \citep{Brunner2022}.

The stacking is done by averaging the background subtracted surface brightness profiles within the same annulus around the group center. Since the background and its noise are the key ingredient of the analysis, we measure the background in two different ways to check for consistency and to properly measure the noise of such measurements. We first measure the background by averaging the background level in an annulus around each group with the same observed area. The distance of the annulus from the group sample is chosen to be 2 Mpc. We exclude all events flagged as point source in the annulus and we subtract the area corresponding to the point source when estimating the surface density. As a second measure, we average the background in a number of random regions equal to the number of the prior groups with a diameter of 10 arcmin. The regions are chosen to avoid any point or extended eROSITA detection. The comparison is done to make sure that the masking procedure is correctly done. The two estimates of the background event density agree very well within 1$\sigma$, confirming that the background level is spatially uniform. Such background density, re-scaled to the area in observed units of the annuli of the surface brightness profile, is subtracted from the observed number of events to obtain the background subtracted surface brightness profile. 

We, thus, perform the stacking analysis by averaging the background subtracted number density of events in physical annuli around the group coordinates. Every luminosity derived from the stacked signal is estimated at the median redshift of the prior sample. By selecting only events with a rest frame energy between 0.5-2 keV at the median redshift of the prior sample, we introduce an error due to the redshift distribution of the sample. To take this into account in the error budget, we use a synthetic representative cluster spectrum and shift it to the redshifts of the prior sample. We estimate the rest frame energy in the rest-frame 0.5-2 keV energy for each prior. We measure the mean and the dispersion of such rest frame energy distribution and compare it to the energy corresponding to the spectrum shifted at the median redshift of the sample. We find that the two estimates are consistent within $1\sigma$, where $\sigma$ is the dispersion over the number of priors. The dispersion is in all cases smaller than $4\%$, as expected, given the relatively narrow redshift window considered in this work.


\begin{figure*}
\includegraphics[width=\textwidth]{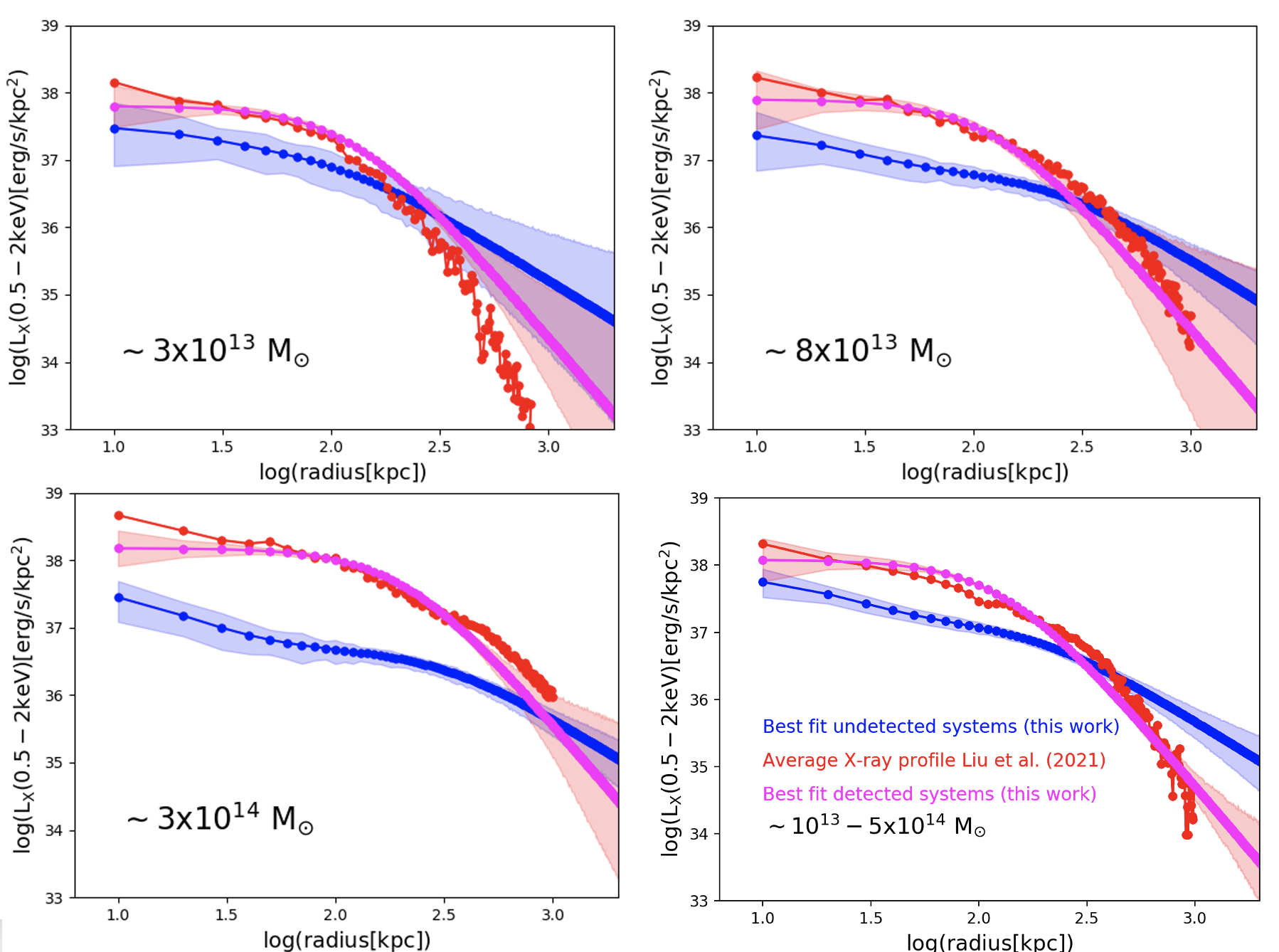}
\caption{The two upper panels and the bottom left panel show the mean X-ray surface brightness of detected and undetected sources in three bins of halo mass, derived from the GAMA velocity dispersion. The bottom right panel shows the same comparison between the detected sample and a mass-matched sub-sample of the undetected systems over the whole available halo mass range. The blue line shows the best fit of the stacked undetected GAMA systems, while the blue-shaded region shows the error of the corresponding observed mean profile. The magenta line shows the best fit of the stacked detected eFEDS systems, and the corresponding shaded region shows the error of the observed mean profile. For comparison, the connected red points show the mean profile obtained by averaging the X-ray surface brightness profile of \citet{LiuAng2022} of the same systems.}
\label{profiles}
\end{figure*}


In order to test the method, we perform the very same analysis on the detections and compare with the mean profiles obtained by averaging the profiles given in \citet{LiuAng2022}. Fig. \ref{profiles} shows the profile obtained with our method (magenta curves) with respect of those of \citep[][red curves]{LiuAng2022} in several halo mass bins. The error is estimated as the dispersion of the average of \citet{LiuAng2022} profiles for the red curve. In the stacking analysis the error of the profile is estimated by boost-trapping over the prior sample. The agreement is remarkably good and always within $1-1.5\sigma$. 

We, thus, conclude that our method is reliable and can provide a robust measure of the mean X-ray surface brightness and luminosity of the undetected sample.

\subsection{The AGN and X-ray binaries contamination}
\label{contamination}
Low luminosities AGN and X-ray binary (XRB) emission might be a source of the contamination of the intra-group medium signal in the stacking analysis. To cope with this issue we use the following approach. 

In order to identify possible low luminosity AGN, we use the optical selection. Thus, we identify among the system members those classified as AGN in the BPT diagram. We use the fluxes of SDSS and AAOmega spectra provided in \citet{Gordon2017}, as suggested also in \citet{Driver2022}. Those represent nearly the totality of the spectra of the group members analyzed here. We classify as AGN all systems above the \citet{Kewley2006} relation. We identify more than 2000 AGN among the entire group GAMA galaxy member population, which correspond to a fraction $\sim 21\%$ of the whole galaxy sample above the considered magnitude limit. This fraction include also LINERs and not only higher luminosity AGN.

In order to estimate the possible contamination, we assign to each AGN, as upper limit, the flux corresponding to the eROSITA point source detection threshold \citep{Salvato2022}. We create a PSF re-scaled to this flux to simulate a point source emission. We use the 10 arcmin radius regions used to estimate the background, one for each undetected system, to distribute over it the eROSITA PSF of the low luminosity AGN with the same spatial distribution as in the parent group. We perform the stacking analysis of these regions, which do not contain any group or point source emission and we measure the contribution of the artificial PSF to the stacking. We do not observe any signal above the $2\sigma$ level and we conclude that the contamination by low luminosity AGN, if any, is marginal with respect to the intra-group medium signal observed in our stacks. 

With respect to the XRB contribution, we apply the same approach of \citep{Chadayammuri2022}. In particular, we express the XRB emission as a function of stellar mass and star formation rate of the group members as proposed in that paper. We renormalize the eROSITA PSF such that the integrated luminosity is equal to that of the expected XRB contribution. We subtract this contribution from each radial bin of the stacks.

\subsection{Results}

We perform the stacking analysis in several halo mass bins by ensuring to have at least 15 systems to stack. Given the halo mass distribution of the undetected sample, the higher the halo mass, the lower the number of systems to stack. 

In all halo mass bins the undetected systems exhibit flatter and more extended surface brightness profiles with respect to the detected systems, although the noise in the system outskirts does not allow a firm conclusion about the system extension (Fig. \ref{profiles}). We point out that the higher the halo mass the larger the difference between the observed surface brightness profiles, with the most massive systems being quite under-luminous within the inner 0.5 Mpc with respect to the detected counterparts. The average difference in X-ray surface brightness in this region increases from $0.5\pm0.1$ dex at $\sim 3\times 10^{13}$ $M_{\odot}$, to $0.63\pm0.12$ dex at $\sim 8\times 10^{13}$ $M_{\odot}$ to $0.84\pm0.13$ dex at $\sim 3\times 10^{14}$ $M_{\odot}$.

In order to check with higher statistics and accuracy whether the X-ray surface brightness profile of the undetected systems is less concentrated and more extended than the detected ones, we perform the stacking over the whole sample without halo mass binning. In order to perform a fair comparison, we select from the undetected sample a sub-sample with the same mass distribution as the detected sample but with higher statistics. The right-bottom panel of Fig. \ref{profiles} shows the comparison between the detected and undetected system profiles. The analysis confirms that the undetected systems exhibit a less concentrated and more extended profile well above the 3$\sigma$ level. 

\subsection{Comparison with eROSITA simulations}
\label{selection_function}
\cite{Comparat2020} create a set of full sky light-cones using the MultiDark and UNIT dark matter-only N-body simulations. They predict the X-ray emission of eROSITA extended systems such as galaxy groups and clusters. Given a set of dark matter halo properties (mass, redshift, ellipticity, offset parameter), they construct an X-ray emissivity profile and image for each halo in the light-cone, by following the eROSITA scanning strategy to produce a list of X-ray photons on the full sky. 

\begin{figure}
\includegraphics[width=0.5\textwidth]{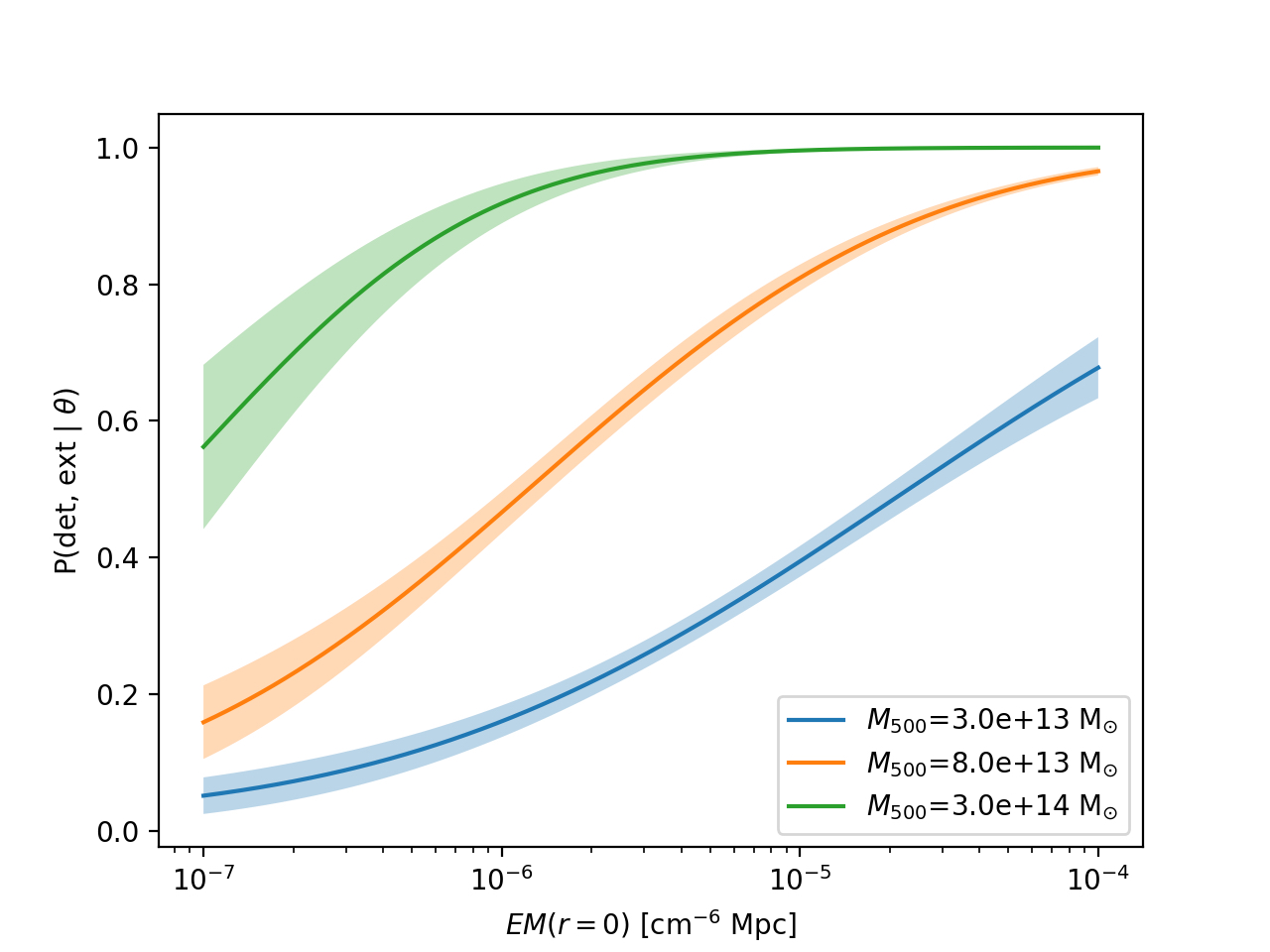}
\caption{Probability of detection by the eFEDS extended object selection technique as a function of the central X-ray emissivity along the line of sight of a system $EM(0)$. $EM(0)$ scales with the system X-ray central brightness, and it is expressed in unit $cm^{-6}$ $Mpc$ for convenience.}
\label{efeds_model}
\end{figure}

By using the selection function models in the eFEDS field based on such end-to-end simulations at $z=0.1$, we derived the detection probability of extended systems in three bins of halo mass corresponding to the halo mass bins of our analysis in section 3.1, as shown in Fig. \ref{efeds_model}.  The three curves are parametrized as a function of the metric $EM(0)$, which represents the central X-ray emissivity along the line of sight of a system. This is derived from the emissivity $EM(r)$ at $r=0$ from eq. 9 in \cite{Comparat2020}.  $EM(0)$ is expressed in unit $cm^{-6}$ $Mpc$ for convenience. $EM(r)$ is proportional to the ratio of the X-ray surface brightness profile and the emissivity of the instrument (see \citet{Comparat2020} for more details). Thus, $EM(0)$ scales with the system X-ray central brightness. The blue and the orange curves, which refer to the group halo masses of $3\times 10^{13}$ and $8\times 10^{13}$ $M_{\odot}$, respectively, clearly show that the eFEDS selection function based on the pipeline settings in \cite{LiuAng2022} favors concentrated objects. This result is based on the choice of balancing sample completeness and contamination by spurious detections. Qualitatively, the predictions are consistent with our results that the largest fractions of groups with a lower central brightness remain undetected in the eFEDS selection. Imposing an effective $EM(0)\sim10^{-7}$ for all objects in the parent sample roughly leads to the observed numbers of undetected halos (blue bars) of Fig. \ref{halomass}. Compared to a nominal $EM(0)=10^{-5.5}$ value, which is usually observed for the X-ray detection samples, this suggests a $\sim 1.5$ difference in the very central X-ray luminosities between undetected and detected systems. Although the quantities cannot be directly compared, this would be roughly consistent with the $\sim 1$ dex luminosity difference in the central parts of detected and undetected systems in Fig. \ref{profiles}.

This would suggest that at fixed halo mass, the X-ray selection is able to capture the fraction of halos with the highest concentration. Since the higher the halo mass, the higher the X-ray luminosity, the concentration threshold leading to the X-ray detection decreases as a function of these quantities. This would naturally explain why we observe a larger discrepancy between the X-ray surface brightness profile at $M_{200}\sim 10^{14}$ $M_{\odot}$ than at lower masses. Indeed, among the poor clusters, which are relatively bright, also the less concentrated systems are likely to be detected, while only the least concentrated ones remain undetected. Thus, the X-ray selection captures most of the objects and only the outliers with a much flatter X-ray surface brightness profile remain below the detection threshold. At lower masses, instead, among the relatively X-ray faint groups, only the most concentrated ones are likely to be detected, while the majority remain undetected due to the low fluxes.

\section{Dynamical analysis}\label{s:dynam}
In this section we perform the dynamical analysis of the detected and undetected systems to understand whether they exhibit consistent dynamical properties or whether and how they differ.

\begin{figure}
\includegraphics[width=0.5\textwidth]{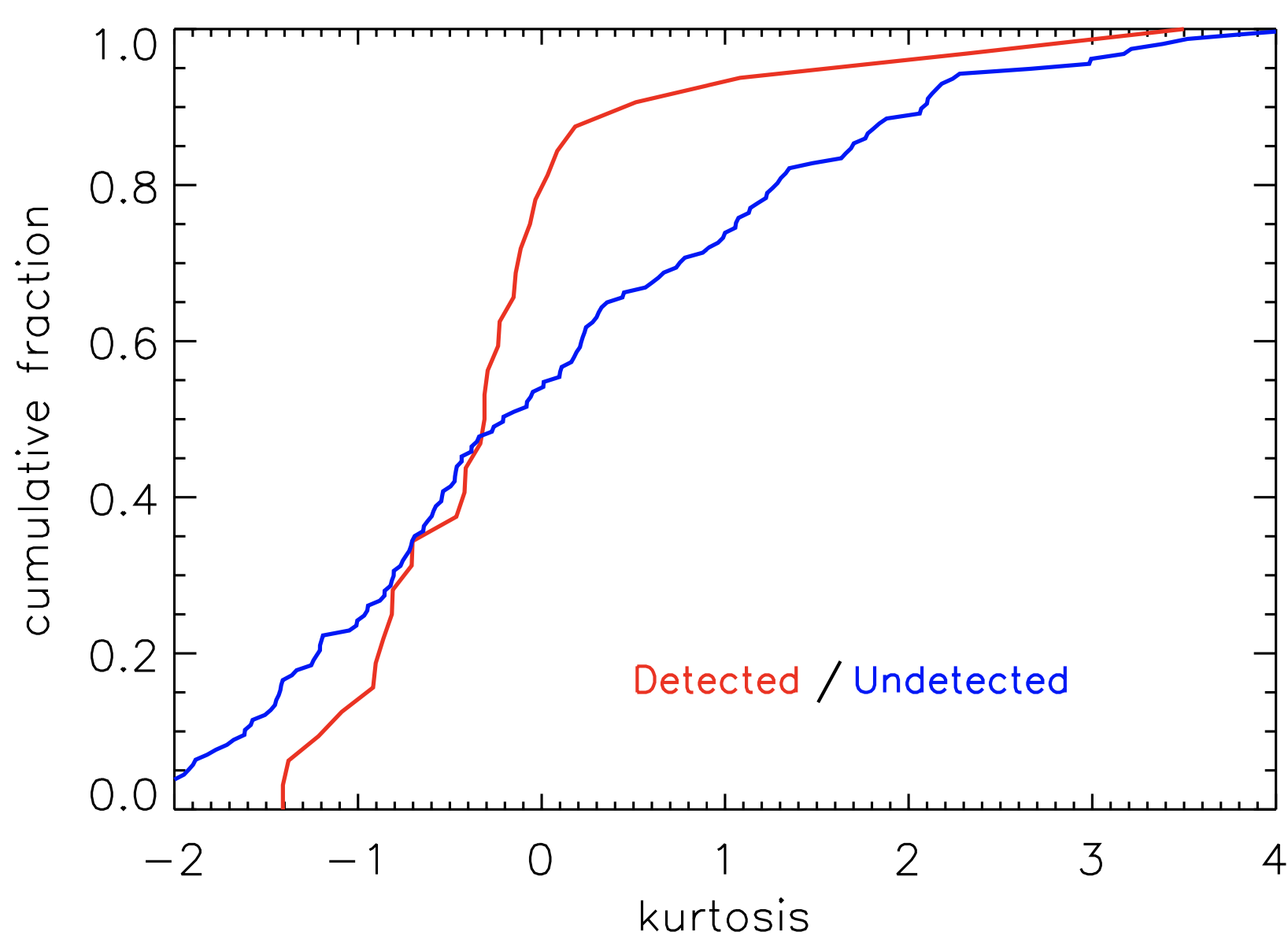}
\caption{Cumulative distribution of the kurtosis of the velocity distributions of X-ray detected (red) and undetected (blue) groups.}
\label{f:kurt}
\end{figure}

For each GAMA system \citet{Robotham2011} provide the kurtosis of the velocity distributions. We display the cumulative distributions of the group velocity distribution kurtosis in Fig.~\ref{f:kurt}, separately for detected, and undetected systems. The kurtosis distribution of the undetected groups is wider than the one of the detected sample. The difference between the two kurtosis distributions is marginally significant, 1.5\%, according to the Kolmogorov-Smirnov test. Undetected groups have velocity distributions that depart more from a Gaussian than detected groups. This suggests they might not be in a fully relaxed dynamical state, but the evidence for this is not statistically significant.

\begin{figure}
\includegraphics[width=0.5\textwidth]{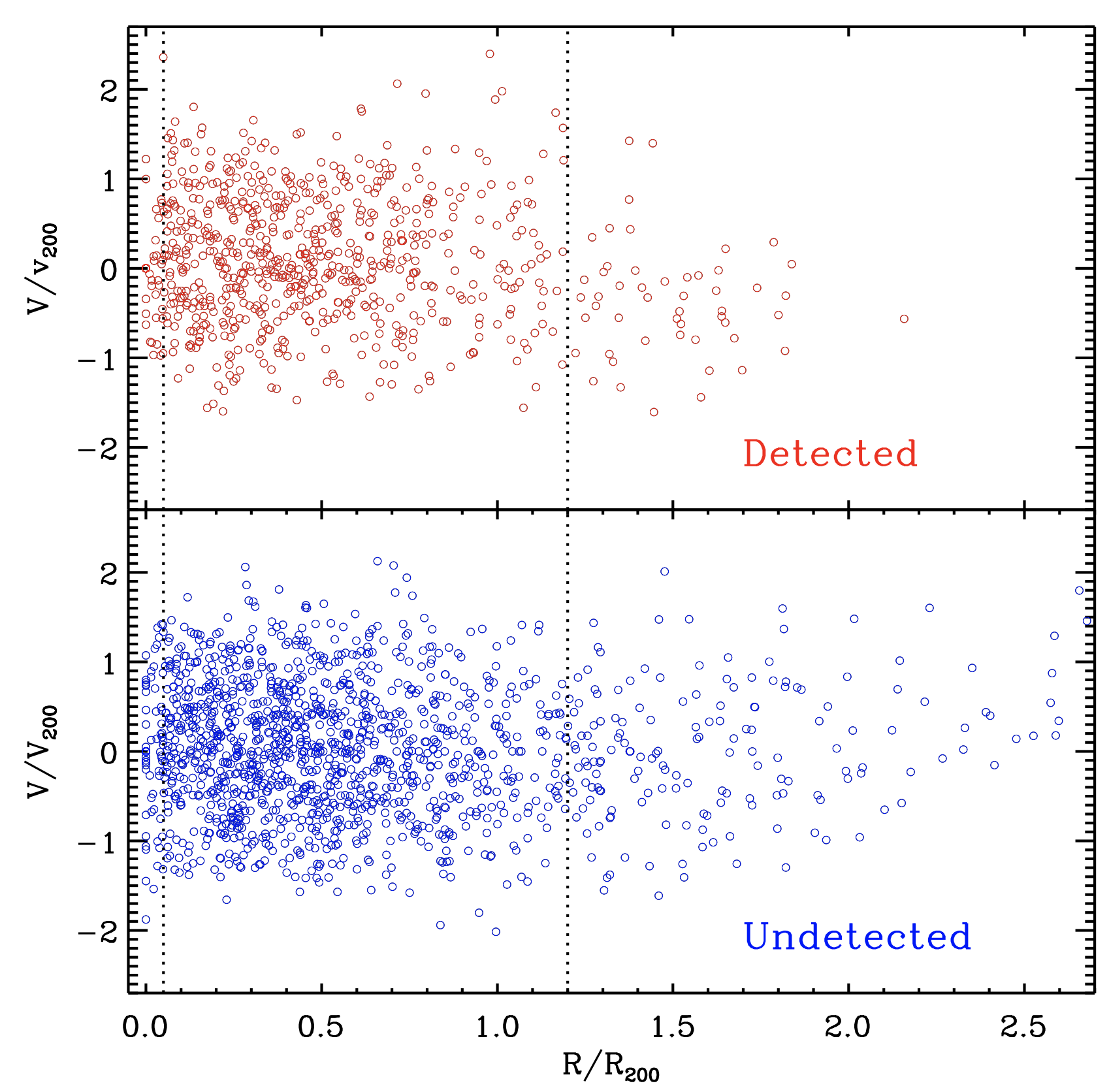}
\caption{Projected-phase-space distributions of the stacked samples of X-ray detected (top panel) and undetected (bottom panel) groups. The dotted line indicates the radial range used in the MAMPOSSt analysis.}
\label{f:RV}
\end{figure}

The above consideration notwithstanding, we proceed with a dynamical analysis of the detected and undetected samples of groups based on the Jeans equation for dynamical equilibrium. The number of group members (multiplicity) is generally too small to allow such dynamical analysis on individual groups. Thus, we proceed with a stacking analysis. Following general practice \citep[e.g.][]{vanderMarel+00,Biviano+21}, we stack the systems in projected phase-space, by normalizing the group-centric projected galaxy distances, $R$, by the group $R_{200}$, and the rest-frame galaxy velocities, $V$, by the group $V_{200}$\footnote{$V_{\Delta}=(G \, M_{\Delta}/R_{\Delta})^{1/2}$}. The quantities $R_{200}$ and $V_{200}$ are derived from the group masses, $M_{200}$, given the adopted cosmology.

In Fig.~\ref{f:RV} we display the projected phase-space distributions of the stacks of the two samples of groups. We examined the velocity distributions of the stacks for deviations from Gaussianity. None of the two distributions have a significant kurtosis. We also look for the tail indices \citep{Beers+91} of the velocity distributions and did not find any significant difference from the Gaussian expectation. Even if some individual group velocity distributions deviate from a Gaussian, the median kurtosis of the group velocity distributions is $\approx 0$ for both samples (see Fig.~\ref{f:kurt}), so it is not unexpected that the stacked distribution is about Gaussian.

\begin{figure}
\includegraphics[width=0.5\textwidth]{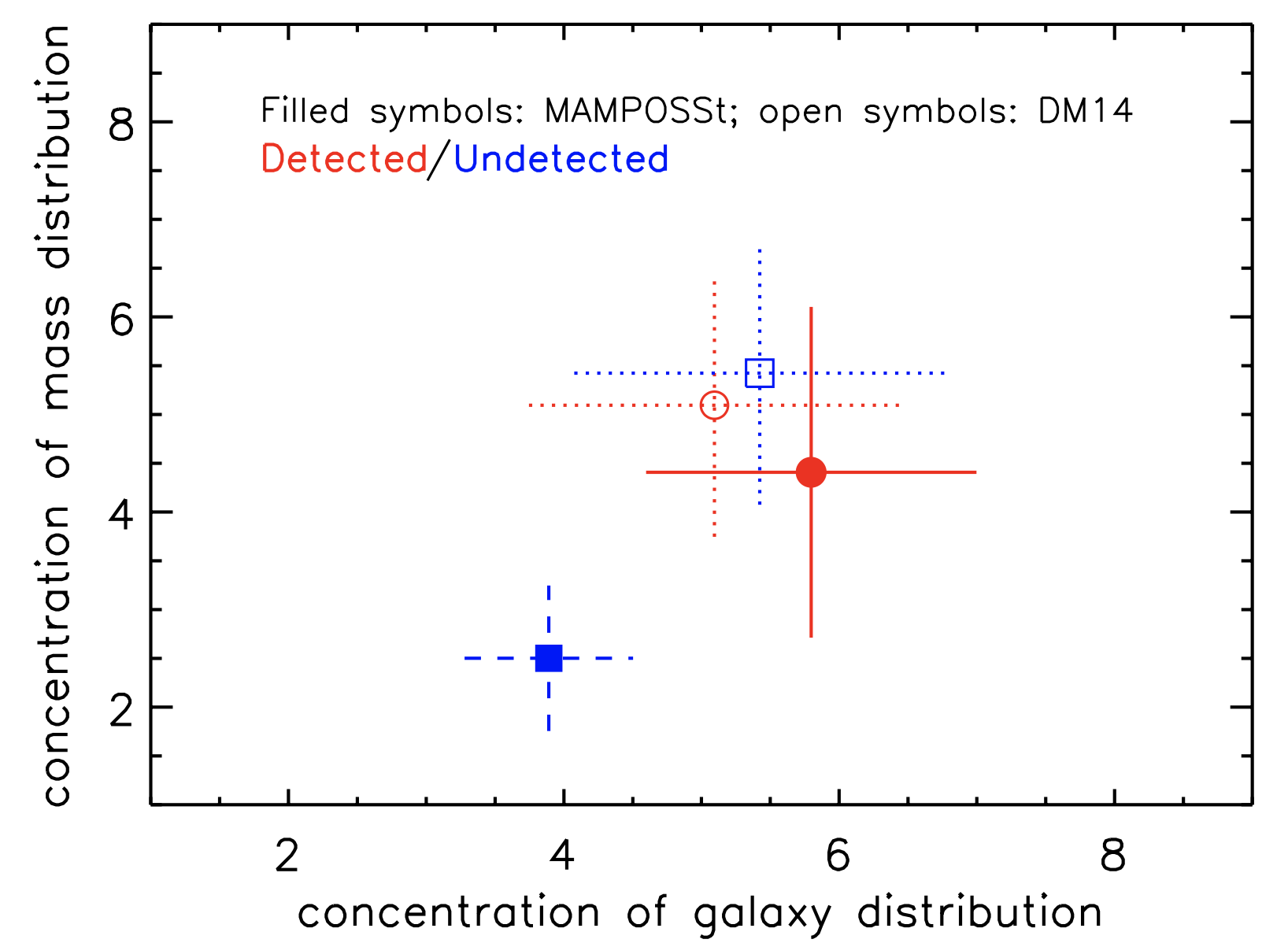}
\caption{Concentration of the distribution of mass, $c_{\rho}$, vs. the concentration of the distribution of group galaxies, $c_g$ for X-ray detected groups (red dot and solid 1-$\sigma$ error bars), and undetected groups (blue square and dashed 1-$\sigma$ error bars). Filled symbols indicate the results of the MAMPOSSt analysis. Open symbols and dotted lines indicate predictions from numerical simulations \citep{DM14} for halos with the same mean $R_{200}$ of the detected and undetected systems (1.0 and 0.8 Mpc, respectively).}
\label{f:conc}
\end{figure}

We then run MAMPOSSt \citep{Mamon2013} to solve the Jeans equation for dynamical equilibrium, separately for each stack. In our MAMPOSSt analysis we only consider galaxies in the \mbox{0.05--1.2 $R_{200}$} radial range, to avoid being affected by the gravitational potential of the BCG at the center, and by dynamical instabilities due to lack of complete virialization in the outskirts. In this radial range there are 613 and 991 galaxies in the detected and, respectively, undetected system stack sample.

We use the code MG-MAMPOSSt of \citet{Pizzuti2022} in GR mode (no modified gravity). We consider the NFW \citep{NFW1997} model for both the galaxy and the mass distributions, characterized by scale radii $R_g$ (galaxies) and $R_{\rho}$ (mass density). We do not consider $R_{200}$ as a free parameter, because both group-centric galaxy distances and galaxy velocities are in normalized units in the stack. The mean values of $R_{200}$ for the detected and undetected stacks are 1.0 and 0.8 Mpc, respectively.

We consider two models for the velocity anisotropy profile $\beta(r)$: the model of \citet{Tiret+07} and the O-M model of \citet{Osipkov79} and \citet{Merritt85-df} \citep[see eq.(10) in][]{BP09}. We find the O-M model to provide a better fit to the velocity distributions of the two samples, so we only provide the results obtained using the O-M model. This model is characterized by an increasing radial velocity anisotropy with radius, with an anisotropy radius $R_{\beta}$ as a free parameter.

We find that the undetected stack sample is characterized by a smaller best-fit $R_{\beta}$ than the detected sample, $0.7 \pm 0.2$ vs. $1.3 \pm 0.6$, respectively. This suggests that the orbits of galaxies in undetected systems are more radially elongated than those in the detected systems. However, the difference is not significant.

In Fig.~\ref{f:conc} we show the results for the concentrations of the galaxy distribution, $c_g \equiv R_{200}/R_g$ (x-axis) and of the mass density distribution, $c_{\rho} \equiv R_{200}/R_{\rho}$. We find $c_g=5.8 \pm 1.2, c_{\rho}=4.1 \pm 1.6$ for the detected sample, and
$c_g=3.9 \pm 0.6, c_{\rho}=2.5 \pm 0.7$ for the undetected samples, respectively. For comparison, we show in the same figure the results from numerical simulations of \citet{DM14} for halos with the $R_{200}$ values of the detected, and, respectively undetected, samples. The numerical predictions for the concentrations are 5.4 and 5.1 for the detected and undetected systems (no distinction between $c_g$ and $c_{\rho}$ is possible since the simulations are dark matter only), with a predicted scatter across the cosmological halos of 1.3.
We find that the detected groups have concentration values in agreement with the numerical predictions, both for the galaxy and the mass density distributions. On the other hand, undetected groups are significantly less concentrated than both detected groups and the predictions from numerical simulations.

\section{Detected versus undetected groups: differences and similarities in global properties}
In this section, we analyze several global properties of the groups and their galaxy population to look for possible differences between the detected and undetected systems, to find the reason  that at fixed halo mass several halos are detected and many others remain invisible to the eROSITA X-ray selection.
We investigate the following properties:
\begin{itemize}
    \item a) the distribution of the gas and the gas mass fraction in the halos
    \item b) the location of the groups within the Cosmic Web from voids, walls, filaments, and nodes;
    \item c) the magnitude gap between the first and second brightest galaxies in the $r$-band, which is considered as a proxy for the system's relative age;
    \item d) the mean color of the Brightest Cluster (or Group) Galaxy (BCG, hereafter); 
    \item e) the mean color of the galaxy population, to test the relative activity and age of the system members;
    \item f) the percentage of optically identified AGN among the system members;
    \item g) the likelihood that the BCG host a radio AGN.
\end{itemize}

\begin{figure*}
\includegraphics[width=\textwidth]{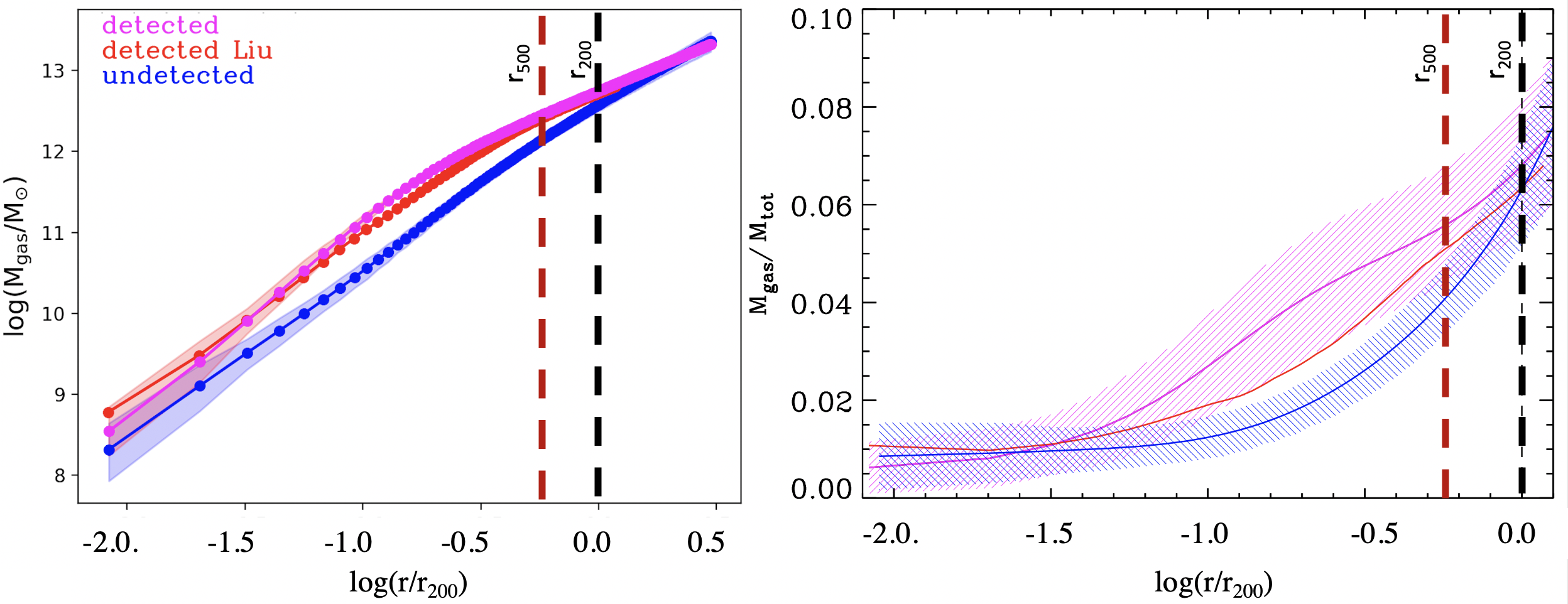}
\caption{{\it{Left panel}}: cumulative gas mass profile as a function of $r/R_{200}$. The blue and magenta lines show the gas mass profile of the stacked undetected GAMA systems and the stacked detected eFEDS systems, respectively. For comparison, the connected red points show the mean profile obtained by averaging the gas mass profile of \citet{LiuAng2022} of the same systems. The shaded regions indicate the uncertainty of the line of the same color. We indicate with a dashed vertical line the location corresponding to $R_{500}$ and $R_{200}$. {\it{Right panel}}: profile of the gas mass fraction as a function of $r/R_{200}$. The color code is the same as in the left panel.}
\label{gas_profile}
\end{figure*}

\begin{figure}
\includegraphics[width=\columnwidth]{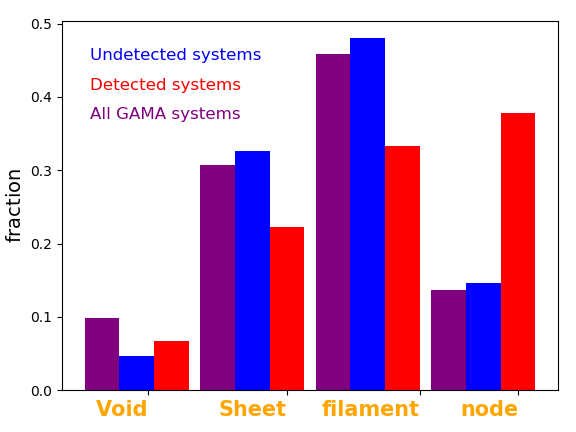}
\caption{Distribution of the X-ray undetected GAMA (blue histogram), X-ray detected eFEDS (red histogram) and the whole GAMA sample with richness higher than 5 (purple histogram) in the 4 classes of the Cosmic Web at $z < 0.2$ identified in the GAMA spectroscopic sample by Eardley et al. (2015)}
\label{cosmic}
\end{figure}

Since the X-ray luminosity and its profile correlates with the gas mass and its distribution within halos, property a) is an obvious aspect to investigate \citep{Lovisari2015}.
Properties b) and c) are chosen to test any possible dependence of the observed difference in halo concentration on the environment, as predicted by the halo assembly bias, and to test for consistency between halo concentration and age \citet{Hahn2009}. The mean color of BCG (d) and satellites (e) will indicate whether there is any difference in the activity and relative age of the  galaxy population \citet{Gozaliasl14}. The last two properties (f and g) have been chosen to test if there is any relation between the observed lower concentration in mass, galaxies, and gas distribution in undetected systems and the overall AGN activity of the galaxy population and the central galaxy \citet{2021Univ....7..142E}. Such analysis is limited to optical and radio AGN activity for obvious reasons. First, we exclude from the stacking analysis of the undetected sample, groups that host an X-ray AGN. Second, given the eROSITA spatial resolution, it is not possible to separate any point-source contribution from the X-ray profile of the detected systems. Thus, we do not know how many X-ray AGN are hosted by the members of the detected sample.

We measure the distribution of the properties listed above and their mean in both the detected and undetected group samples. To measure the significance of any difference and to take into account possible dependencies on the halo mass distribution, we extract 10000 random sub-samples among the undetected systems, each with the same number of groups and halo mass distribution of the detected sample. 

\begin{figure*}
\includegraphics[width=\textwidth]{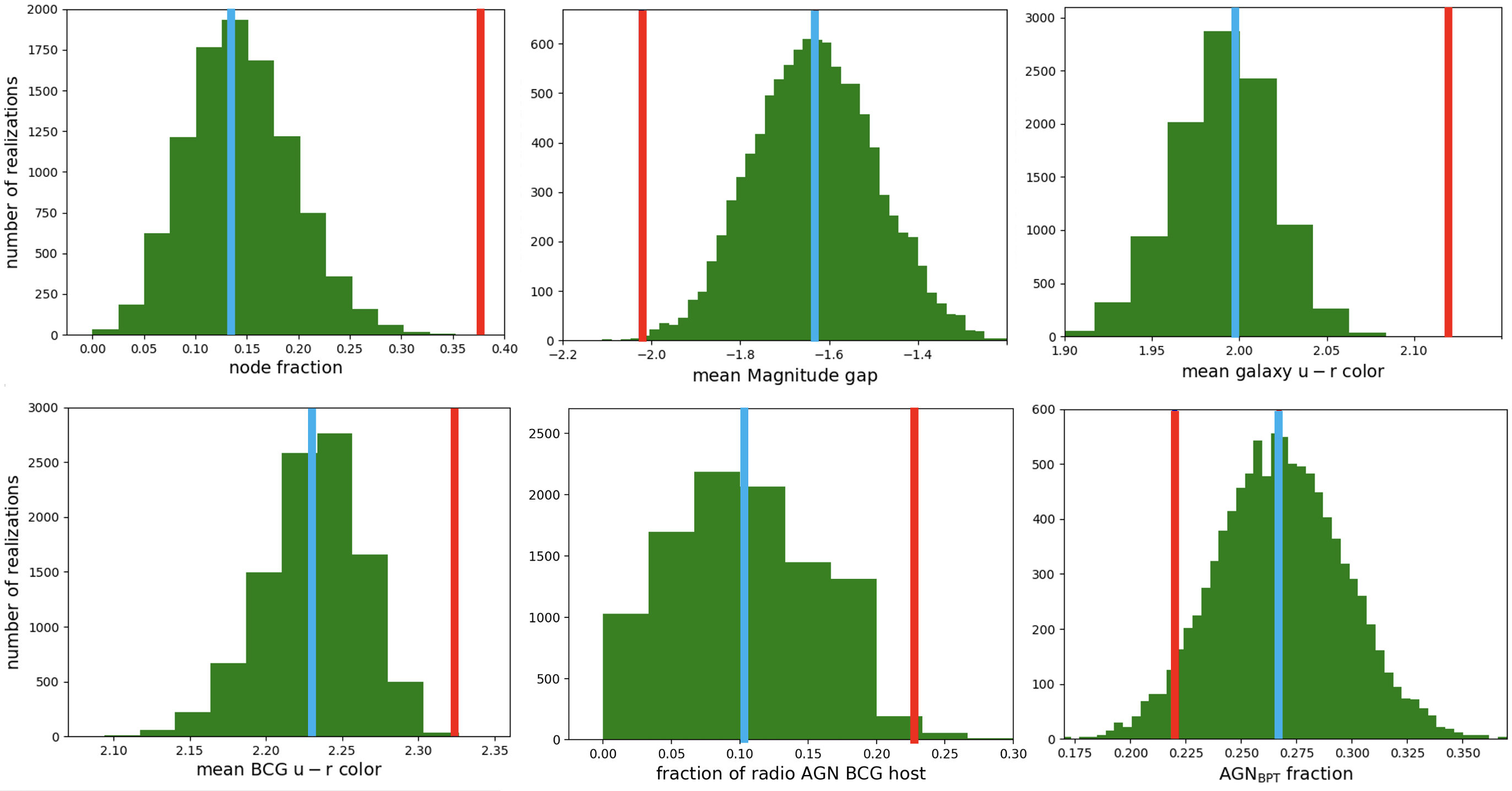}
\caption{Comparison of the mean properties of the detected systems (red line in all panels) and the mean properties of 10000 halo mass-matched samples randomly extracted from the undetected system sample. Each panel shows the corresponding mean value (cyan line) and the distribution in the 10000 random extractions (green histogram). The upper panels from left to right show the comparison of {\it{i)}} the fraction of systems localized in the nodes of the Cosmic web, {\it{ii)}} the mean magnitude gap between the first and the second brightest galaxies in the r band, and {\it{iii)}} the mean $u-r$ rest frame color of the galaxy population per group. The bottom panels from left to right show: {\it{iv)}} the mean $u-r$ rest frame color of the BCG, {\it{v)}} the mean fraction of optical AGN hosts per group and {\it{vi)}} the probability that the BCG host a radio AGN.}
\label{histo}
\end{figure*}

\subsection{The gas mass distribution}
The gas mass profile of the eFEDS detection has been estimated as in \citet{LiuAng2022}. For consistency, we derive the gas mass fraction from the stacked X-ray surface brightness profile of the undetected systems with the same approach used for the eFEDS detections. We assign to the stacks the temperature corresponding to the mean mass ($M_{500}$) of the stacked groups, according to the $T-M_{500}$ relation of \citet{Lovisari2015}. We refer to \citet{LiuAng2022} for the details of the procedure. We show here the results for the subsample of undetected systems mass-matched to the eFEDS sample. 

The left panel of Fig. \ref{gas_profile} shows the comparison of the cumulative gas mass profile for the detected and undetected systems. While the mass distribution differs within the core region, accordingly to the X-ray surface brightness profile, the total gas mass is nearly the same within $R_{200}$ and it differs by a factor of $\sim1.5$ within $R_{500}$. We show also in the right panel of Fig. \ref{gas_profile} the gas mass fraction profile, obtained by dividing the gas mass profile by the best fit NFW mass profile estimated with MAMPOSSt. Since also the NFW mass profile of the undetected systems is less concentrated than for the detected systems, the gas mass fraction profiles differ by less than 1.5$\sigma$ and provide the same gas fraction within 1$\sigma$ within $R_{500}$ and $R_{200}$.

\subsection{The Cosmic Web}

For locating the groups within the Cosmic Web, we use the catalog of \citet{CW2015}. This classifies the entire GAMA area over a smoothing length of 4 to 10 Mpc by providing 4 Cosmic Web categories: voids, sheets, filaments, and nodes. The classification is based on the Cosmic Web reconstruction derived from the spatial and redshift distribution of the highly complete GAMA galaxy spectroscopic sample. A grid of cells of RA, Dec, and redshift, with an associated Cosmic Web category, is provided over the entire GAMA area and up to $z=0.2$. By associating galaxies and groups to a cell it is possible to identify the corresponding Cosmic Web category. We do this for the detected, undetected, and parent samples. The location of detected and undetected systems differ significantly. As shown in Fig. \ref{cosmic} the distribution of the detected (red histograms) and undetected systems (blue histograms) across the different cosmic web components tend to differ significantly. A percentage of 38\% of the detected systems are located at the nodes of the Cosmic web, while only 13\% of the undetected groups are in the same cosmic Web component. The majority of them favor filaments or sheets. To check if this difference is due to different halo mass distributions, we compare the percentage of detected groups in nodes with the percentages estimated in the 10000 mass-matched sub-samples of undetected groups. As shown in the upper left panel of Fig. \ref{histo}, the percentage of detected groups in nodes is at least 5 $\sigma$ higher than the mean percentage of the undetected groups.

\subsection{The magnitude gap and the activity of the BCG and the group galaxy population}
Similarly, we measure the magnitude gap in the detected and the 10000 extractions of the undetected detected sample. The magnitude gap of the detected groups is 3$\sigma$ larger than for the undetected groups, suggesting that the detected systems tend to be older than the undetected ones. This is consistent with the findings of \citet{Giles2022} based on the XXM-XXL survey data. This leads also to another interesting aspect. Since we compare halo mass-matched samples, the larger magnitude gap indicate also that at fixed halo mass, the stellar mass of the BCG is, on average, larger in the detected systems than in the undetected ones. This is due to the fact that the absolute $r_{mag}$ correlates with the stellar mass although with a relatively large scatter ($\sim 0.2$ dex in the GAMA sample). Such aspect is confirmed by the fact that detected systems tend to populate the upper envelope of the $M_{BCG}/M_{halo}-M_{halo}$ relation, as shown in the left panel of Fig. \ref{scaling}. The mean distance from the $M_{BCG}/M_{halo}-M_{halo}$ relation of \citet{Berhoozi2019} 
is consistent with 0 for the GAMA parent sample, consistently with the \citet{2007ApJ...671..153Y} and the \citet{Tempel2017} samples. The detected systems, instead, are slightly displaced with respect to the relation in the upper envelope with a mean distance of $0.16\pm0.05$ dex.

The relatively younger dynamical age of the undetected systems is reflected in a higher activity of the BCG and the galaxy population. We measure the level of activity through the rest frame $u-r$ color, which has been largely used to separate active from quiescent galaxies in particular in the SDSS galaxy sample \citep{Strateva2001}. Both the color of the BCG and, as expected from galactic conformity \citep{Weinmann2006}, the color of the rest of the galaxy population with $r_{mag} < -20$, are bluer than the color of the BCG and, respectively, of the galaxy population, of the detected groups. The comparison shown in Fig. \ref{histo} indicates that the statistical significance of the difference is higher than 3$\sigma$. 

\subsection{The level of nuclear activity of BCG and galaxy members}
As the last test, we measure the mean percentage of optically selected AGN among the group members brighter than $r_{mag}=-20$ and within the inner 0.5 Mpc, and the fraction of BCGs hosting a radio AGN. The optical AGN are selected as explained in section \ref{contamination}

The central bottom panel in Fig. \ref{histo}  shows the comparison between the mean fraction of AGN hosts per group in the detected sample (red line) and the mean (blue line) and distribution (green histogram) of the same value in the 10000 extractions of the undetected sample mass-matched to the detected sample. Undetected systems show on average a slightly higher fraction of AGN hosts per group. However, the significance of the effect is only at the 2$\sigma$ level.

The radio AGN are classified on the basis of the LOFAR 144MHz data available for this field \citep{Pasini2022}. Firstly, the LOFAR detections are filtered with a $SNR > 5$. They have then matched to the DESI Legacy DR9 survey \citep{Dey2019} using a Bayesian multiwavelength cross-matching algorithm called NWAY \citep{Salvato2018}. This allows using not only distance priors but also magnitude priors in the cross-matching procedure. This is important when matching radio to optical, as radio sources tend to be found in redder galaxies \citep[e.g.][]{Williams2019}. Details of the cross-matching can be found in Igo et al. (in prep), including details on selecting a subsample of sources that optimizes the purity and completeness, as well as distinguishing complex and compact radio emitters.

Among the LOFAR radio sources, AGN are identified in two ways. To be conservative, we apply a cut in radio Power. In the local Universe, the radio luminosity function of star-forming galaxies and AGN are separated at $\log(P_{1.4GHz}) \sim 23$ $WHz^{-1}$. By assuming a radio spectral index 0.7, such a limit becomes $\log(P_{150MHz}) \sim 23.7$ $WHz^{-1}$ in the LOFAR band. By using this method, we identify 95 radio AGN, of which 36 are in satellites and 59 in BCGs.
Alternatively, we follow also the approach of \citet{DelVecchio2021} by identifying radio AGN as those which exhibit an extra radio emission with respect to their SFR, namely as the sources above the SFR-radio power correlation. This method is usually applied by using the far-infrared luminosity radio-power correlation. However, the H-ATLAS data in the GAMA field is too shallow for this purpose. We thus use the SFR estimated with MagPhys for each galaxy, using multi-wavelength observations covering from far UV ($\sim 1500 \, \AA$) to far infrared \citep[$\sim 500 \, \mu$m][]{Robotham2020, Bellstedt2021}. 
By choosing systems 3$\sigma$ above the correlation, we identify 308 AGN, of which 103 are in satellites and 205 in BCGs.

Both selection methods reveal that radio AGN are more common among BCG than satellites, as expected due to the known mass segregation of the radio AGN population \citep[see for instance][]{Sabater2019, Best2005}. In addition, the number of radio AGN identified in both ways among satellites is negligible with respect to the satellite population of the detected and undetected systems. For these reasons, we only consider the radio AGN fractions among BCGs, and not among the satellites. As shown in the bottom right panel of Fig. \ref{histo}, BCGs in detected systems have a probability 3 times higher ($35\%$) to host a radio AGN than BCGs in undetected systems. The significance of the effect is at the 5$\sigma$ level as shown by the comparison with respect to the 10000 mass-matched extractions from the undetected system sample. This is confirmed by both radio AGN selection methods, although the absolute value of the probability differs, being smaller for the more conservative power cut selection. 

This effect is consistent with the findings highlighted above. Indeed, the detected systems exhibit a larger magnitude gap, indicating that the BCG is brighter and consistently more massive than in the mass-matched undetected group sub-samples. Since radio AGN tend to segregate in the most massive galaxies, this might cause the observed larger probability that the BCG of detected systems host a radio AGN.

\section{The $L_X-M_{halo}$ relation}
\begin{figure}
\includegraphics[width=\columnwidth]{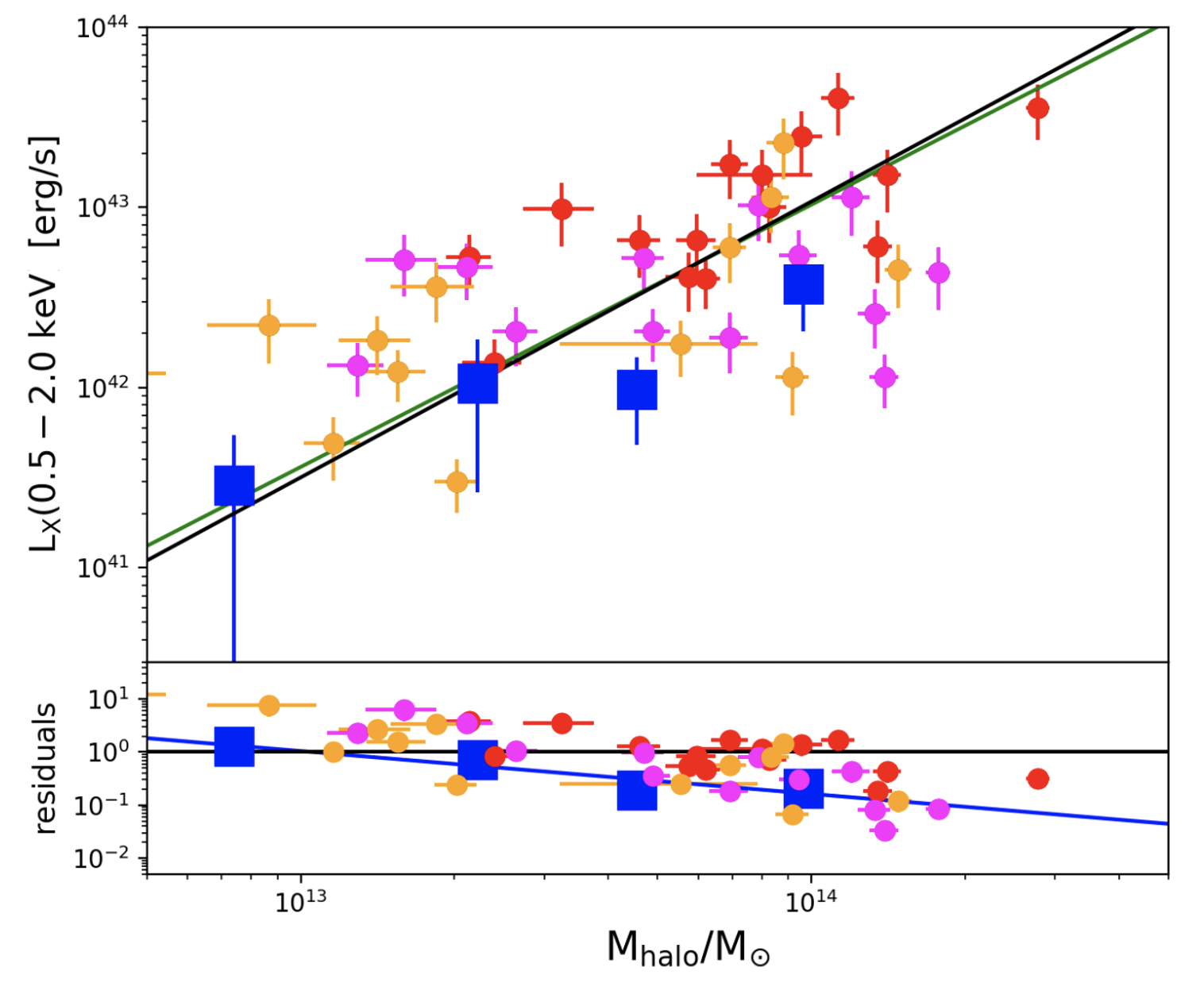}
\caption{Relation between the halo mass ($M_{500}$) and the X-ray luminosity estimated with $r_{500}$ in the 0.5-2.0 keV (upper panel). The detected systems are indicated by the points. These are color-coded as a function of their location in the Cosmic Web: nodes (red), filaments (magenta), and sheets and void (orange). The blue squares show the X-ray luminosity derived from the stacks in different halo mass bins. The black solid line shows the $L_X-M_{500}$ relation in the 0.5-2 keV band of \citet{pratt09}, while the green line is the one of \citet{Chiu2022} based on eFEDS groups and clusters. The two relations are nearly identical. The bottom panel shows the residuals with respect to the relation of \citet{pratt09} The blue solid line shows the linear regression best fit to the stacks residuals.}
\label{lxmhalo}
\end{figure}

\begin{figure}
\includegraphics[width=\columnwidth]{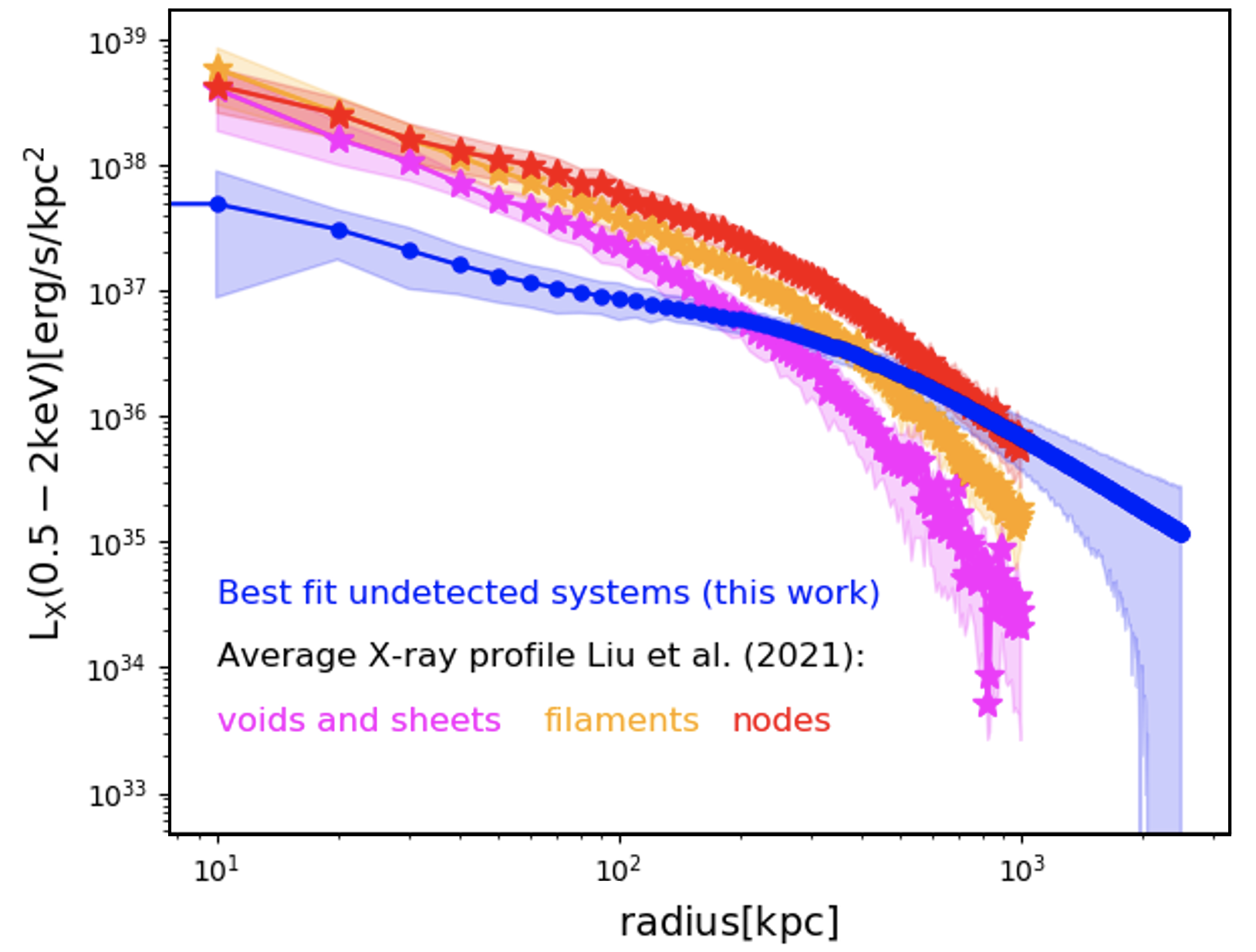}
\caption{The panel shows the mean X-ray surface brightness of the detected systems provided by \citet{LiuAng2022} in three Cosmic Web components: voids and sheets (magenta profile), filaments (orange profile), and nodes (red profiles). The blue line region shows the mean stacked profile of the undetected systems with a similar halo mass distribution as for the detected systems.  The shaded regions for all profiles represent the 1$\sigma$ error retrieved from the bootstrapping analysis. The median of the mass distribution is the same for all components.}
\label{cosmic_web}
\end{figure}

We analyze here the location of detected and undetected systems in the
$L_X-M_{halo}$ diagram. We measure the halo mass and the X-ray luminosity within $R_{500}$. For the systems, detected and undetected, we use the estimate of $R_{500}$ and $M_{500}$ as described in Section \ref{dataset}. 
For the detected systems, we measure the X-ray luminosity from the surface brightness profile of \citet{LiuAng2022} by integrating up to our measure of the $R_{500}$. For the stacks, we use the median $R_{500}$ of the corresponding prior sample to integrate the average surface brightness profile, while the mass is estimated as the median $M_{500}$ of the prior systems (see Section \ref{dataset}).

The $L_X-M_{halo}$ relation is shown in Fig. \ref{lxmhalo}. The points in the figure show the detected systems, while the big squares are the stacks of the undetected systems. Since one of the most significant differences between the detected and undetected systems is the location within the Cosmic Web, we color code also the detected systems as a function of their location within nodes (red points), filaments (orange points), and sheets and voids (magenta points). The detected systems located in the nodes of the cosmic web are those with the smallest scatter around the $0.5-2.0$ keV energy-band $L_X-M_{halo}$ relation of \citet[][black solid line]{pratt09} and \citet[][green solid line]{Chiu2022}. Above $2\times 10^{13}$ $M_{\odot}$, the detected systems located outside the nodes of the Cosmic Web scatter below the relation, on average by $0.41\pm0.11$ dex in X-ray luminosity. In this halo mass range the stacks lie significantly below the relation of the REXCESS clusters with a scatter of $ 0.7\pm 0.12$ dex. 

In order to understand if, above $2\times 10^{13}$ $M_{\odot}$, the distance from the REXCESS relation depends on the system's location in the Cosmic Web, we analyze the mean X-ray surface brightness profile of the detected and undetected systems as a function of the Cosmic Web location. Fig. \ref{cosmic_web} shows the mean profiles of the detected systems located in different cosmic Web components, as obtained by averaging the profiles provided by \citet{LiuAng2022}. The inner panel in the figure shows the $M_{500}$ distributions of the detected systems per Cosmic Web components. We point out that the distributions are consistent, and show a median mass of $8.3\pm0.4\times$, $8.6\pm 0.5\times$ and $8.7\pm 0.4\times 10^{13}$ $M_{\odot}$, in sheets, filaments, and nodes, respectively. Thus, any difference in the observed surface brightness profile can not be ascribed to mass segregation effects. For comparison, we show also the mean surface brightness profile of the undetected systems (blue curve) with the same mass distribution of the detected systems. We can not divide the undetected systems per cosmic Web component because the statistics is not sufficient in this restricted halo mass range to obtain a sufficiently significant stacked signal. However, as pointed out in the previous section, the undetected systems are mostly ($\sim 89\%$) located outside of the nodes, with a high frequency (47\%) in filaments. Thus, we consider them as representative of the cosmic web components. The shaded regions around each profile show the $1\sigma$ error, which is obtained by bootstrapping over the sample. We observe a clear decrease in the mean X-ray emission from nodes, filaments to sheets. In particular, at fixed halo mass, the X-ray emission of a halo is higher in the outskirts in crowded regions such as the nodes, and it decreases progressively from filaments to sheets and voids, making the profile of the halos in nodes less concentrated that the one in the other cosmic web components. On the contrary, the X-ray surface brightness profile of the undetected systems is much flatter, as already found in the previous section (blue profile in Fig. \ref{cosmic_web}). We tend to exclude ascribing the flat profile of the stacked profile to a high level of contamination of spurious optically selected groups. We point out that the inclusion of such contaminants in the undetected sample should affect mainly the SNR of the stacks in the eROSITA data rather than their overall profile. In addition, particularly large contamination should heavily affect also the stacks of the velocity dispersion distribution, which, instead, is consistent with a Gaussian. 

We argue that since the undetected systems outnumber the detected ones outside the nodes, their profile represents the average X-ray surface brightness profile of halos in filaments and sheets. Since the X-ray emission depends on the ${\rho^2}_{gas}$, where $\rho_{gas}$ is the density of the hot gas, at such low X-ray luminosities, the X-ray selection would be able to capture only the tail of the most concentrated halos per cosmic web component. Indeed, as shown in Fig. \ref{detectability}, if we measure the detection probability, as the ratio between the detected and parent sample as a function of the Cosmic Web component, the probability is the highest for the nodes ($\sim 76\%$) and it decreases to $25-30\%$ in the less crowded Cosmic Web components. Since the eFEDS depth is slightly higher than the nominal depth of the future eRASS at the completion, we conclude that, in the considered halo mass range, the eRASS X-ray selection will be biased towards highly concentrated halos and will naturally favor the detection of halos in nodes. This is consistent with the current studies of the eFEDS and eRASS selection function as shown in section \ref{selection_function}.

Moving to lower X-ray luminosities and halo masses, below $2\times 10^{13}$ $M_{\odot}$, in the $L_X-M_{halo}$ relation we observe that the stacks of the undetected systems are consistent with the REXCESS relation. However, we point out that the optically selected sample used here is able to capture only half of the halo population, when compared with the predictions of the Millennium light-cones \citep[][see also Fig. \ref{halomass}]{Smith2017}.  The X-ray selected sample, instead, captures only a negligible percentage. This is due to the fact that the cut in richness applied to the optically selected sample might bias the estimate as it captures the richest groups of the underlying halo population. Such a significant bias limits the interpretation of the results in this halo mass range and requires a different technique to select halos. Thus, we will tackle this issue in a separate paper.

\section{Discussion \& Conclusions}

In this section, we provide a possible interpretation of our results. Overall we find that the X-ray undetected systems exhibit a higher probability to reside outside of the nodes of the cosmic web. Their X-ray surface brightness profile is less concentrated than the detected systems. Consistently, the dynamical analysis indicates that the distribution of galaxies and the distribution of mass are both less concentrated in undetected systems than in the detected ones, and they are also below the theoretical predictions. Consequently, the gas mass fraction distribution does not vary significantly between detected and undetected systems. This might suggest that the undetected groups are in a post-merger phase, as suggested by their wider distribution of velocity kurtosis values compared to detected groups. Mergers with a small impact radius can cause the distributions of dark matter, galaxies, and intra-group gas, to become less concentrated, as in the well-studied case of the cluster Abell 315 \citep{Biviano2017}. In this case, the question remains about how many of these halos reached virialization.

In addition, the magnitude gap on average is smaller for the undetected systems pointing to a younger formation epoch with respect to the detected counterparts. This is consistent with the lower concentration derived independently from the dynamical and the X-ray analysis. Indeed, N-body simulations show that the higher the concentration the earlier the formation epoch of the system \citep{Wechsler2002}. The later assembly time is also reflected in a relatively younger and more active galaxy population as proven by the bluer average color of the BCG and of the galaxy population with respect to the detected systems. 

\begin{figure}
\includegraphics[width=\columnwidth]{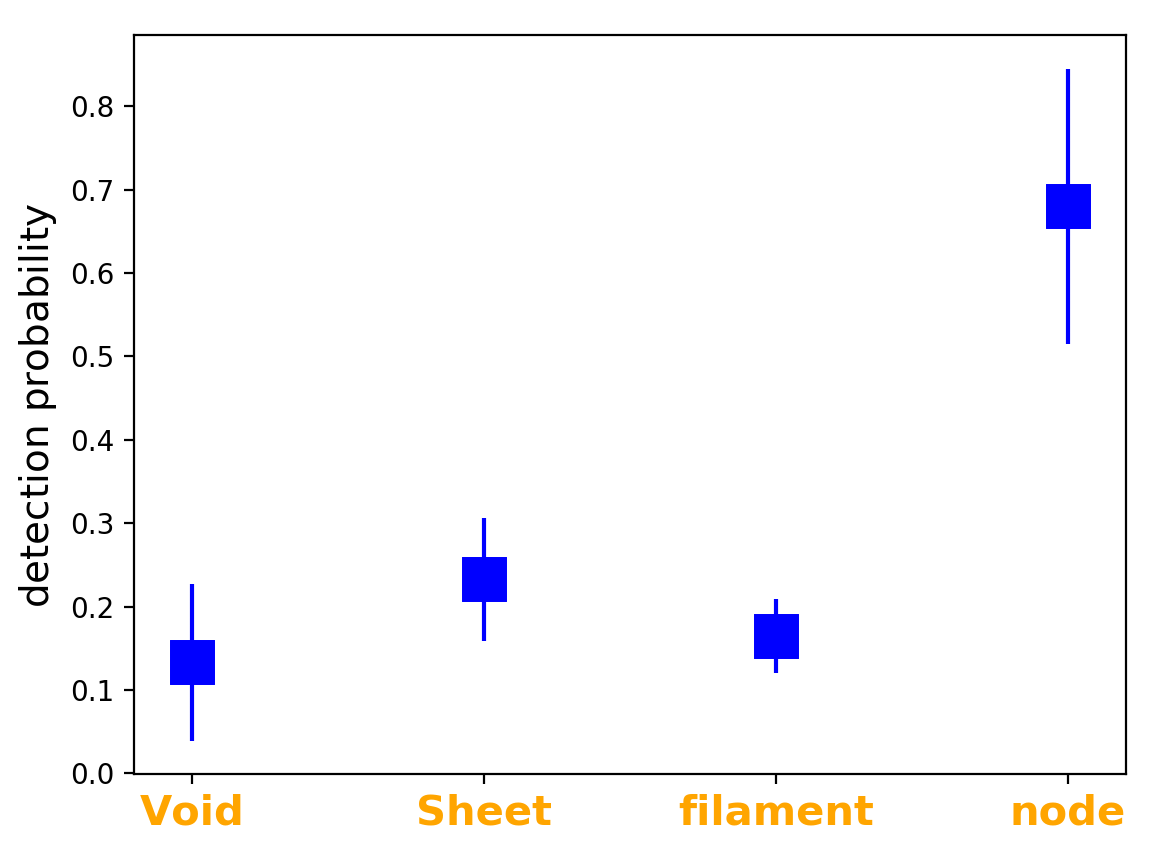}
\caption{Detection probability of halos above $2\times 10^{13}$ $M_{\odot}$ per Cosmic Web component in the eROSITA soft band at the eFEDS depth. The probability is estimated as the fraction of eFEDS detection over the number of the underlying halo population per Cosmic Web component, represented by our sub-sample of optically selected GAMA systems. }
\label{detectability}
\end{figure}

We also observe that among the detected systems, above halo masses of $\sim 2\times 10^{13}$ $M_{\odot}$,  those located in the nodes exhibit a higher X-ray surface brightness in the outskirts, with a profile less concentrated than those located in filaments and in the least crowded regions as sheets and voids. This leads to a larger deviation from the $L_X-M_{500}$ relation of \citet{pratt09} and \citet{Chiu2022} towards lower X-ray luminosities as a function of the Cosmic Web component. 

Conversely, we do not observe any strong relation between the less concentrated mass, galaxy, and density profile of the undetected systems and the AGN activity. The effect of outflows and radio jets, as feedback generated from the central black hole accretion, would in principle be able to lead to a lower gas concentration in the center by removing part of the gas towards the system outskirts. However, there is no evidence in simulations that AGN feedback might have a direct effect on the distribution of collisionless tracers of the potential such as galaxies or dark matter. \citep{Peirani2019} suggest, by using HORIZON-AGN, that by displacing baryonic mass from the center to outside, the feedback reduces the depth of the central gravitational potential well, and can cause dark matter to follow this change by gravitational interaction. However, such results are limited to the galactic scale and are used to show that, without AGN feedback, the simulations would produce too compact red galaxies with respect to observations. Thus, while the feedback might have an influence on the dark matter and star distribution on small galactic scale (a few to tens of kpc), there is no indication that it could affect the matter and galaxy distribution on Mpc scale, as observed here.

In addition, we point out that we observe only a marginally higher fraction of optical AGN in the undetected systems with respect to the detected ones, at the $2\sigma$ significance level. Instead, it is more interesting to note, that the frequency of radio AGN activity in the BCG is significantly lower (4$\sigma$ significance level) in the undetected systems than in the detected ones. Only 1 over 10 undetected systems exhibit radio-activity in the central galaxy, while 1 over 4 detected systems have a BCG hosting a radio AGN. If anything, this would point to a lower probability of feedback effects in the undetected systems analyzed here than in the detected ones. At the same time, this suggests that radio AGN activity may be more likely triggered at the center of more concentrated halos; however, an in-depth investigation of the connection between halo properties and AGN triggering is beyond the scope of this work. We point out that our results are at odds with the predictions of \citet{Ragagnin22} based on the Magneticum simulation. Indeed, they conclude that older galaxy clusters tend to be gas-poor and possess a low X-ray surface brightness, due to the efficient removal of the gas from the central region from the AGN feedback over the halo lifetime. As a consequence, under-luminous clusters tend to exhibit lower gas fractions. We can not exclude that, despite the current low AGN activity of the central galaxy of the undetected systems, those groups might have been heavily affected by feedback in the past. However, the observed gas mass fraction profile, the dynamical analysis of the systems, the magnitude gap analysis, and the bluer colors of BCG and of the overall galaxy population of the undetected systems do not support such a scenario.

We argue here that the observed differences and similarities between detected and undetected systems might be ascribable to the Cosmic Web influence. The fact that the Cosmic Web affects the halo properties is well known. Recent studies indicate that the cosmic web environment at relatively small scales - of the order of a few virial radii - plays an important role in the assembly bias by directly affecting the halo formation epoch \citep{Hahn2009}, the mass accretion rate \citep{Fakhouri_Ma2010,Musso2018}, the internal velocity dispersion structure \citep{Borzyszkowski2017}, and the halo concentration \citep{Paranjape2018}. In particular, these studies have revealed an intimate connection between the nature of assembly bias and the immediate environment of a halo, e.g. whether or not the halo lives in a cosmic filament or in a node \citep[see also][]{Shi2015}. While this is not unexpected \citep{Bond1996,Sheth2001,Musso2012, Castorina2013}, the specific role of the non-linear cosmic web in this matter is still highly debated and not completely understood. It is still unclear whether the local environmental density, e.g. over 4 Mpc scale, might be driving the correlation rather than the tidal forces related to the larger scale environment of the filaments. Indeed, translating environmental over-densities and clustering amplitude into Cosmic Web components is not as straightforward as it might appear. As shown in \citet{Paranjape2018}, the over-density distributions of halos in nodes, filaments, and sheets, in particular, overlap largely. They show that halos in filaments tend to cluster more strongly than halos in nodes, in particular in the vicinity of the filament intersections. Such halos, which might have undergone a higher merger activity may plausibly be less centrally concentrated than halos of the same mass which have more quiescent accretion histories. Lower mass halos, in the same crowded Cosmic Web component, would, instead, suffer from the tidal forces of the more massive halos, which would act in stopping the mass accretion rate and increasing the central density of the halos \citep{Ramakrishnan2019, Paranjape2018, Paranjape2017}.


Since the range of halo masses considered in this paper is limited to the massive groups and clusters, above $10^{13}$ $M_{\odot}$, we compare our results with the predictions of the halo assembly bias above $m*$ (see the introduction), which according to \citet{Ramakrishnan2019} and \citep{Paranjape2018} is at $2-3\times 10^{13}$ $M_{\odot}$. The undetected systems in eFEDS are mainly located in filaments and sheets, where they dominate in number over the detected systems. This can be seen in Fig. \ref{detectability}, where we show the percentage of detected systems over the underlying halo population, represented by our parent sample. Thus, it is logical to conclude that they represent the average halo population in such cosmic web components. It follows that, according to our results, halos in these Cosmic Web components exhibit a lower concentration and a younger age with respect to the halos in the nodes, which are mostly sampled by the detected systems (see Fig. \ref{detectability}). These exhibit a slightly higher concentration and earlier formation epoch. These results would be consistent with the results of \citet{Ramakrishnan2019}, \citet{Paranjape2018}, and  \citet{Paranjape2017}, although the lower statistics do not allow to distinguish and stack the undetected systems per Cosmic Web component.

In addition, by measuring the connectivity, which is the number of filaments a halo is connected to, \citet{Gouin2021} find in the Illustris TNG simulation that the location within the Cosmic environment might strongly affect the dynamical status of halos. In particular, groups at $z\sim0$ in the same halo mass range as analyzed here, appear to be separated into two populations: one with high connectivity with filaments and still in a formation process, and one of highly concentrated and old groups disconnected from the surrounding environment. The former population represents the majority. This could be, in principle, in line with our findings. However, we point out that the different definitions of the cosmic environment and the use of different algorithms to identify its components might lead to systematic differences. In this respect, observations and simulations should be analyzed consistently with the same methods to assess a proper comparison. 

In terms of selection biases, our results would suggest that the X-ray selection, in the halo mass range studied here, is highly biased toward the selection of the concentrated and old halos. This would imply that eRASS will mostly detect halos in the nodes of the Cosmic Web and those at the tail towards high values of concentration in the halo concentration distribution of the respective cosmic web component. The much larger statistics offered by the combination of eRASS and the SDSS galaxy group samples, with respect to eFEDS and GAMA, will allow to constraint with higher accuracy the dependence of the halo concentration on the cosmic web components and study the nature of the low mass halos without biases.

\section*{ACKNOWLEDGEMENTS}
We thank the referee for her/his very useful comments.

This project has received funding from the European Research Council (ERC) under the European Union’s Horizon Europe research and innovation programme ERC CoG (Grant agreement No. 101045437).

This work is based on data from eROSITA, the soft X-ray instrument aboard SRG, a joint Russian-German science mission supported by the Russian Space Agency (Roskosmos), in the interests of the Russian Academy of Sciences represented by its Space Research Institute (IKI), and the Deutsches Zentrum für Luft- und Raumfahrt (DLR). The SRG spacecraft was built by Lavochkin Association (NPOL) and its subcontractors and is operated by NPOL with support from the Max Planck Institute for Extraterrestrial Physics (MPE).

The development and construction of the eROSITA X-ray instrument was led by MPE, with contributions from the Dr. Karl Remeis Observatory Bamberg \& ECAP (FAU Erlangen-Nuernberg), the University of Hamburg Observatory, the Leibniz Institute for Astrophysics Potsdam (AIP), and the Institute for Astronomy and Astrophysics of the University of Tübingen, with the support of DLR and the Max Planck Society. The Argelander Institute for Astronomy of the University of Bonn and the Ludwig Maximilians Universität Munich also participated in the science preparation for eROSITA.

The eROSITA data shown here were processed using the eSASS software system developed by the German eROSITA consortium.

\section*{Data Availability}

All data used in this paper are publicly available through the GAMA data release \citep{Driver2022} and the eFEDS data release \citet{LiuAng2022}.



\bibliographystyle{mnras}
\bibliography{laura1} 

\begin{thebibliography}{}
\makeatletter
\relax
\def\mn@urlcharsother{\let\do\@makeother \do\$\do\&\do\#\do\^\do\_\do\%\do\~}
\def\mn@doi{\begingroup\mn@urlcharsother \@ifnextchar [ {\mn@doi@} {\mn@doi@[]}}
\def\mn@doi@[#1]#2{\def\@tempa{#1}\ifx\@tempa\@empty \href {http://dx.doi.org/#2} {doi:#2}\else \href {http://dx.doi.org/#2} {#1}\fi \endgroup}
\def\mn@eprint#1#2{\mn@eprint@#1:#2::\@nil}
\def\mn@eprint@arXiv#1{\href {http://arxiv.org/abs/#1} {{\tt arXiv:#1}}}
\def\mn@eprint@dblp#1{\href {http://dblp.uni-trier.de/rec/bibtex/#1.xml} {dblp:#1}}
\def\mn@eprint@#1:#2:#3:#4\@nil{\def\@tempa {#1}\def\@tempb {#2}\def\@tempc {#3}\ifx \@tempc \@empty \let \@tempc \@tempb \let \@tempb \@tempa \fi \ifx \@tempb \@empty \def\@tempb {arXiv}\fi \@ifundefined {mn@eprint@\@tempb}{\@tempb:\@tempc}{\expandafter \expandafter \csname mn@eprint@\@tempb\endcsname \expandafter{\@tempc}}}

\bibitem[\protect\citeauthoryear{{Anderson}, {Gaspari}, {White}, {Wang}  \& {Dai}}{{Anderson} et~al.}{2015}]{Anderson2015}
{Anderson} M.~E.,  {Gaspari} M.,  {White} S. D.~M.,  {Wang} W.,   {Dai} X.,  2015, \mn@doi [\mnras] {10.1093/mnras/stv437}, \href {https://ui.adsabs.harvard.edu/abs/2015MNRAS.449.3806A} {449, 3806}

\bibitem[\protect\citeauthoryear{{Beers}, {Gebhardt}, {Forman}, {Huchra}  \& {Jones}}{{Beers} et~al.}{1991}]{Beers+91}
{Beers} T.~C.,  {Gebhardt} K.,  {Forman} W.,  {Huchra} J.~P.,   {Jones} C.,  1991, \aj, 102, 1581

\bibitem[\protect\citeauthoryear{{Behroozi}, {Wechsler}, {Hearin}  \& {Conroy}}{{Behroozi} et~al.}{2019}]{Berhoozi2019}
{Behroozi} P.,  {Wechsler} R.~H.,  {Hearin} A.~P.,   {Conroy} C.,  2019, \mn@doi [\mnras] {10.1093/mnras/stz1182}, \href {https://ui.adsabs.harvard.edu/abs/2019MNRAS.488.3143B} {488, 3143}

\bibitem[\protect\citeauthoryear{{Bellstedt} et~al.,}{{Bellstedt} et~al.}{2021}]{Bellstedt2021}
{Bellstedt} S.,  et~al., 2021, \mn@doi [\mnras] {10.1093/mnras/stab550}, \href {https://ui.adsabs.harvard.edu/abs/2021MNRAS.503.3309B} {503, 3309}

\bibitem[\protect\citeauthoryear{{Best}, {Kauffmann}, {Heckman}  \& {Ivezi{\'c}}}{{Best} et~al.}{2005}]{Best2005}
{Best} P.~N.,  {Kauffmann} G.,  {Heckman} T.~M.,   {Ivezi{\'c}} {\v{Z}}.,  2005, \mn@doi [\mnras] {10.1111/j.1365-2966.2005.09283.x}, \href {https://ui.adsabs.harvard.edu/abs/2005MNRAS.362....9B} {362, 9}

\bibitem[\protect\citeauthoryear{{Biviano} \& {Poggianti}}{{Biviano} \& {Poggianti}}{2009}]{BP09}
{Biviano} A.,  {Poggianti} B.~M.,  2009, \mn@doi [\aap] {10.1051/0004-6361/200911757}, 501, 419

\bibitem[\protect\citeauthoryear{{Biviano}, {Popesso}, {Dietrich}, {Zhang}, {Erfanianfar}, {Romaniello}  \& {Sartoris}}{{Biviano} et~al.}{2017}]{Biviano2017}
{Biviano} A.,  {Popesso} P.,  {Dietrich} J.~P.,  {Zhang} Y.~Y.,  {Erfanianfar} G.,  {Romaniello} M.,   {Sartoris} B.,  2017, \mn@doi [\aap] {10.1051/0004-6361/201629471}, \href {https://ui.adsabs.harvard.edu/abs/2017A&A...602A..20B} {602, A20}

\bibitem[\protect\citeauthoryear{{Biviano} et~al.,}{{Biviano} et~al.}{2021}]{Biviano+21}
{Biviano} A.,  et~al., 2021, \mn@doi [\aap] {10.1051/0004-6361/202140564}, \href {https://ui.adsabs.harvard.edu/abs/2021A&A...650A.105B} {650, A105}

\bibitem[\protect\citeauthoryear{{Blanton} et~al.,}{{Blanton} et~al.}{2017}]{blanton2017}
{Blanton} M.~R.,  et~al., 2017, \mn@doi [\aj] {10.3847/1538-3881/aa7567}, \href {https://ui.adsabs.harvard.edu/abs/2017AJ....154...28B} {154, 28}

\bibitem[\protect\citeauthoryear{{Bond} \& {Myers}}{{Bond} \& {Myers}}{1996}]{Bond1996}
{Bond} J.~R.,  {Myers} S.~T.,  1996, \mn@doi [\apjs] {10.1086/192269}, \href {https://ui.adsabs.harvard.edu/abs/1996ApJS..103...63B} {103, 63}

\bibitem[\protect\citeauthoryear{{Borzyszkowski}, {Porciani}, {Romano-D{\'\i}az}  \& {Garaldi}}{{Borzyszkowski} et~al.}{2017}]{Borzyszkowski2017}
{Borzyszkowski} M.,  {Porciani} C.,  {Romano-D{\'\i}az} E.,   {Garaldi} E.,  2017, \mn@doi [\mnras] {10.1093/mnras/stx873}, \href {https://ui.adsabs.harvard.edu/abs/2017MNRAS.469..594B} {469, 594}

\bibitem[\protect\citeauthoryear{{Brunner} et~al.,}{{Brunner} et~al.}{2022}]{Brunner2022}
{Brunner} H.,  et~al., 2022, \mn@doi [\aap] {10.1051/0004-6361/202141266}, \href {https://ui.adsabs.harvard.edu/abs/2022A&A...661A...1B} {661, A1}

\bibitem[\protect\citeauthoryear{{Bulbul} et~al.,}{{Bulbul} et~al.}{2022}]{Bulbul2022}
{Bulbul} E.,  et~al., 2022, \mn@doi [\aap] {10.1051/0004-6361/202142460}, \href {https://ui.adsabs.harvard.edu/abs/2022A&A...661A..10B} {661, A10}

\bibitem[\protect\citeauthoryear{{Castorina} \& {Sheth}}{{Castorina} \& {Sheth}}{2013}]{Castorina2013}
{Castorina} E.,  {Sheth} R.~K.,  2013, \mn@doi [\mnras] {10.1093/mnras/stt824}, \href {https://ui.adsabs.harvard.edu/abs/2013MNRAS.433.1529C} {433, 1529}

\bibitem[\protect\citeauthoryear{{Chadayammuri}, {Bogdan}, {Oppenheimer}, {Kraft}, {Forman}  \& {Jones}}{{Chadayammuri} et~al.}{2022}]{Chadayammuri2022}
{Chadayammuri} U.,  {Bogdan} A.,  {Oppenheimer} B.,  {Kraft} R.,  {Forman} W.,   {Jones} C.,  2022, arXiv e-prints, \href {https://ui.adsabs.harvard.edu/abs/2022arXiv220301356C} {p. arXiv:2203.01356}

\bibitem[\protect\citeauthoryear{{Chiu} et~al.,}{{Chiu} et~al.}{2022}]{Chiu2022}
{Chiu} I.~N.,  et~al., 2022, \mn@doi [\aap] {10.1051/0004-6361/202141755}, \href {https://ui.adsabs.harvard.edu/abs/2022A&A...661A..11C} {661, A11}

\bibitem[\protect\citeauthoryear{{Clerc} et~al.,}{{Clerc} et~al.}{2018}]{Clerc2018}
{Clerc} N.,  et~al., 2018, \mn@doi [\aap] {10.1051/0004-6361/201732119}, \href {https://ui.adsabs.harvard.edu/abs/2018A&A...617A..92C} {617, A92}

\bibitem[\protect\citeauthoryear{{Comparat} et~al.,}{{Comparat} et~al.}{2020}]{Comparat2020}
{Comparat} J.,  et~al., 2020, \mn@doi [The Open Journal of Astrophysics] {10.21105/astro.2008.08404}, \href {https://ui.adsabs.harvard.edu/abs/2020OJAp....3E..13C} {3, 13}

\bibitem[\protect\citeauthoryear{{Comparat} et~al.,}{{Comparat} et~al.}{2022}]{Comparat2022}
{Comparat} J.,  et~al., 2022, arXiv e-prints, \href {https://ui.adsabs.harvard.edu/abs/2022arXiv220105169C} {p. arXiv:2201.05169}

\bibitem[\protect\citeauthoryear{{Crossett} et~al.,}{{Crossett} et~al.}{2022}]{Crossett2022}
{Crossett} J.~P.,  et~al., 2022, arXiv e-prints, \href {https://ui.adsabs.harvard.edu/abs/2022arXiv220314200C} {p. arXiv:2203.14200}

\bibitem[\protect\citeauthoryear{{Dalal}, {White}, {Bond}  \& {Shirokov}}{{Dalal} et~al.}{2008}]{dalal_etal08}
{Dalal} N.,  {White} M.,  {Bond} J.~R.,   {Shirokov} A.,  2008, \mn@doi [\apj] {10.1086/591512}, \href {http://adsabs.harvard.edu/abs/2008ApJ...687...12D} {687, 12}

\bibitem[\protect\citeauthoryear{{Delvecchio} et~al.,}{{Delvecchio} et~al.}{2021}]{DelVecchio2021}
{Delvecchio} I.,  et~al., 2021, \mn@doi [\aap] {10.1051/0004-6361/202039647}, \href {https://ui.adsabs.harvard.edu/abs/2021A&A...647A.123D} {647, A123}

\bibitem[\protect\citeauthoryear{{Desjacques}}{{Desjacques}}{2008}]{Desjacques2008}
{Desjacques} V.,  2008, \mn@doi [\mnras] {10.1111/j.1365-2966.2008.13420.x}, \href {https://ui.adsabs.harvard.edu/abs/2008MNRAS.388..638D} {388, 638}

\bibitem[\protect\citeauthoryear{{Dey} et~al.,}{{Dey} et~al.}{2019}]{Dey2019}
{Dey} A.,  et~al., 2019, \mn@doi [\aj] {10.3847/1538-3881/ab089d}, \href {https://ui.adsabs.harvard.edu/abs/2019AJ....157..168D} {157, 168}

\bibitem[\protect\citeauthoryear{{Driver} et~al.,}{{Driver} et~al.}{2022}]{Driver2022}
{Driver} S.~P.,  et~al., 2022, \mn@doi [\mnras] {10.1093/mnras/stac472}, \href {https://ui.adsabs.harvard.edu/abs/2022MNRAS.513..439D} {513, 439}

\bibitem[\protect\citeauthoryear{{Dutton} \& {Macci{\`o}}}{{Dutton} \& {Macci{\`o}}}{2014}]{DM14}
{Dutton} A.~A.,  {Macci{\`o}} A.~V.,  2014, \mn@doi [\mnras] {10.1093/mnras/stu742}, \href {http://adsabs.harvard.edu/abs/2014MNRAS.441.3359D} {441, 3359}

\bibitem[\protect\citeauthoryear{{Eardley} et~al.,}{{Eardley} et~al.}{2015}]{CW2015}
{Eardley} E.,  et~al., 2015, \mn@doi [\mnras] {10.1093/mnras/stv237}, \href {https://ui.adsabs.harvard.edu/abs/2015MNRAS.448.3665E} {448, 3665}

\bibitem[\protect\citeauthoryear{{Eckert}, {Gaspari}, {Gastaldello}, {Le Brun}  \& {O'Sullivan}}{{Eckert} et~al.}{2021}]{2021Univ....7..142E}
{Eckert} D.,  {Gaspari} M.,  {Gastaldello} F.,  {Le Brun} A. M.~C.,   {O'Sullivan} E.,  2021, \mn@doi [Universe] {10.3390/universe7050142}, \href {https://ui.adsabs.harvard.edu/abs/2021Univ....7..142E} {7, 142}

\bibitem[\protect\citeauthoryear{{Efstathiou}, {Frenk}, {White}  \& {Davis}}{{Efstathiou} et~al.}{1988}]{Efstathiou1988}
{Efstathiou} G.,  {Frenk} C.~S.,  {White} S.~D.~M.,   {Davis} M.,  1988, \mn@doi [\mnras] {10.1093/mnras/235.3.715}, \href {http://ads.nao.ac.jp/abs/1988MNRAS.235..715E} {235, 715}

\bibitem[\protect\citeauthoryear{{Fakhouri} \& {Ma}}{{Fakhouri} \& {Ma}}{2010}]{Fakhouri_Ma2010}
{Fakhouri} O.,  {Ma} C.-P.,  2010, \mn@doi [\mnras] {10.1111/j.1365-2966.2009.15844.x}, \href {https://ui.adsabs.harvard.edu/abs/2010MNRAS.401.2245F} {401, 2245}

\bibitem[\protect\citeauthoryear{{Faltenbacher} \& {White}}{{Faltenbacher} \& {White}}{2010}]{faltenbacher_white10}
{Faltenbacher} A.,  {White} S.~D.~M.,  2010, \mn@doi [\apj] {10.1088/0004-637X/708/1/469}, \href {http://adsabs.harvard.edu/abs/2010ApJ...708..469F} {708, 469}

\bibitem[\protect\citeauthoryear{{Gao} \& {White}}{{Gao} \& {White}}{2007}]{gao_white07}
{Gao} L.,  {White} S.~D.~M.,  2007, \mn@doi [\mnras] {10.1111/j.1745-3933.2007.00292.x}, \href {http://adsabs.harvard.edu/abs/2007MNRAS.377L...5G} {377, L5}

\bibitem[\protect\citeauthoryear{{Gao}, {Springel}  \& {White}}{{Gao} et~al.}{2005}]{gao_etal05}
{Gao} L.,  {Springel} V.,   {White} S.~D.~M.,  2005, \mn@doi [\mnras] {10.1111/j.1745-3933.2005.00084.x}, \href {http://adsabs.harvard.edu/abs/2005MNRAS.363L..66G} {363, L66}

\bibitem[\protect\citeauthoryear{{Gastaldello}, {Simionescu}, {Mernier}, {Biffi}, {Gaspari}, {Sato}  \& {Matsushita}}{{Gastaldello} et~al.}{2021}]{2021Univ....7..208G}
{Gastaldello} F.,  {Simionescu} A.,  {Mernier} F.,  {Biffi} V.,  {Gaspari} M.,  {Sato} K.,   {Matsushita} K.,  2021, \mn@doi [Universe] {10.3390/universe7070208}, \href {https://ui.adsabs.harvard.edu/abs/2021Univ....7..208G} {7, 208}

\bibitem[\protect\citeauthoryear{{Giles} et~al.,}{{Giles} et~al.}{2022}]{Giles2022}
{Giles} P.~A.,  et~al., 2022, \mn@doi [\mnras] {10.1093/mnras/stab3626}, \href {https://ui.adsabs.harvard.edu/abs/2022MNRAS.511.1227G} {511, 1227}

\bibitem[\protect\citeauthoryear{{Gordon} et~al.,}{{Gordon} et~al.}{2017}]{Gordon2017}
{Gordon} Y.~A.,  et~al., 2017, \mn@doi [\mnras] {10.1093/mnras/stw2925}, \href {https://ui.adsabs.harvard.edu/abs/2017MNRAS.465.2671G} {465, 2671}

\bibitem[\protect\citeauthoryear{{Gouin}, {Bonnaire}  \& {Aghanim}}{{Gouin} et~al.}{2021}]{Gouin2021}
{Gouin} C.,  {Bonnaire} T.,   {Aghanim} N.,  2021, \mn@doi [\aap] {10.1051/0004-6361/202140327}, \href {https://ui.adsabs.harvard.edu/abs/2021A&A...651A..56G} {651, A56}

\bibitem[\protect\citeauthoryear{{Gozaliasl} et~al.,}{{Gozaliasl} et~al.}{2014}]{Gozaliasl14}
{Gozaliasl} G.,  et~al., 2014, \mn@doi [\aap] {10.1051/0004-6361/201322459}, \href {https://ui.adsabs.harvard.edu/abs/2014A&A...566A.140G} {566, A140}

\bibitem[\protect\citeauthoryear{{Hahn}, {Porciani}, {Dekel}  \& {Carollo}}{{Hahn} et~al.}{2009}]{Hahn2009}
{Hahn} O.,  {Porciani} C.,  {Dekel} A.,   {Carollo} C.~M.,  2009, \mn@doi [\mnras] {10.1111/j.1365-2966.2009.15271.x}, \href {https://ui.adsabs.harvard.edu/abs/2009MNRAS.398.1742H} {398, 1742}

\bibitem[\protect\citeauthoryear{{Kaiser}}{{Kaiser}}{1984}]{Kaiser1984}
{Kaiser} N.,  1984, \mn@doi [\apjl] {10.1086/184341}, \href {http://ads.nao.ac.jp/abs/1984ApJ...284L...9K} {284, L9}

\bibitem[\protect\citeauthoryear{{Kewley}, {Groves}, {Kauffmann}  \& {Heckman}}{{Kewley} et~al.}{2006}]{Kewley2006}
{Kewley} L.~J.,  {Groves} B.,  {Kauffmann} G.,   {Heckman} T.,  2006, \mn@doi [\mnras] {10.1111/j.1365-2966.2006.10859.x}, \href {https://ui.adsabs.harvard.edu/abs/2006MNRAS.372..961K} {372, 961}

\bibitem[\protect\citeauthoryear{{Klein} et~al.,}{{Klein} et~al.}{2022}]{Klein2022}
{Klein} M.,  et~al., 2022, \mn@doi [\aap] {10.1051/0004-6361/202141123}, \href {https://ui.adsabs.harvard.edu/abs/2022A&A...661A...4K} {661, A4}

\bibitem[\protect\citeauthoryear{{Le Brun}, {McCarthy}, {Schaye}  \& {Ponman}}{{Le Brun} et~al.}{2014}]{LeBrun2014}
{Le Brun} A. M.~C.,  {McCarthy} I.~G.,  {Schaye} J.,   {Ponman} T.~J.,  2014, \mn@doi [\mnras] {10.1093/mnras/stu608}, \href {https://ui.adsabs.harvard.edu/abs/2014MNRAS.441.1270L} {441, 1270}

\bibitem[\protect\citeauthoryear{{Lim}, {Mo}, {Lu}, {Wang}  \& {Yang}}{{Lim} et~al.}{2017}]{Lim2017}
{Lim} S.~H.,  {Mo} H.~J.,  {Lu} Y.,  {Wang} H.,   {Yang} X.,  2017, \mn@doi [\mnras] {10.1093/mnras/stx1462}, \href {https://ui.adsabs.harvard.edu/abs/2017MNRAS.470.2982L} {470, 2982}

\bibitem[\protect\citeauthoryear{{Liu} et~al.,}{{Liu} et~al.}{2022a}]{LiuAng2022}
{Liu} A.,  et~al., 2022a, \mn@doi [\aap] {10.1051/0004-6361/202141120}, \href {https://ui.adsabs.harvard.edu/abs/2022A&A...661A...2L} {661, A2}

\bibitem[\protect\citeauthoryear{{Liu} et~al.,}{{Liu} et~al.}{2022b}]{LiuTeng2022}
{Liu} T.,  et~al., 2022b, \mn@doi [\aap] {10.1051/0004-6361/202141643}, \href {https://ui.adsabs.harvard.edu/abs/2022A&A...661A...5L} {661, A5}

\bibitem[\protect\citeauthoryear{{Lovisari} \& {Ettori}}{{Lovisari} \& {Ettori}}{2021}]{2021Univ....7..254L}
{Lovisari} L.,  {Ettori} S.,  2021, \mn@doi [Universe] {10.3390/universe7080254}, \href {https://ui.adsabs.harvard.edu/abs/2021Univ....7..254L} {7, 254}

\bibitem[\protect\citeauthoryear{{Lovisari}, {Reiprich}  \& {Schellenberger}}{{Lovisari} et~al.}{2015}]{Lovisari2015}
{Lovisari} L.,  {Reiprich} T.~H.,   {Schellenberger} G.,  2015, \mn@doi [\aap] {10.1051/0004-6361/201423954}, \href {https://ui.adsabs.harvard.edu/abs/2015A&A...573A.118L} {573, A118}

\bibitem[\protect\citeauthoryear{{Lovisari}, {Ettori}, {Gaspari}  \& {Giles}}{{Lovisari} et~al.}{2021}]{2021Univ....7..139L}
{Lovisari} L.,  {Ettori} S.,  {Gaspari} M.,   {Giles} P.~A.,  2021, \mn@doi [Universe] {10.3390/universe7050139}, \href {https://ui.adsabs.harvard.edu/abs/2021Univ....7..139L} {7, 139}

\bibitem[\protect\citeauthoryear{{Mamon}, {Biviano}  \& {Bou{\'e}}}{{Mamon} et~al.}{2013}]{Mamon2013}
{Mamon} G.~A.,  {Biviano} A.,   {Bou{\'e}} G.,  2013, \mn@doi [\mnras] {10.1093/mnras/sts565}, \href {https://ui.adsabs.harvard.edu/abs/2013MNRAS.429.3079M} {429, 3079}

\bibitem[\protect\citeauthoryear{{Mantz} et~al.,}{{Mantz} et~al.}{2015}]{Mantz2015}
{Mantz} A.~B.,  et~al., 2015, \mn@doi [\mnras] {10.1093/mnras/stu2096}, \href {https://ui.adsabs.harvard.edu/abs/2015MNRAS.446.2205M} {446, 2205}

\bibitem[\protect\citeauthoryear{{Merritt}}{{Merritt}}{1985}]{Merritt85-df}
{Merritt} D.,  1985, \mnras, 214, 25P

\bibitem[\protect\citeauthoryear{{Mo} \& {White}}{{Mo} \& {White}}{1996}]{MoWhite1996}
{Mo} H.~J.,  {White} S.~D.~M.,  1996, \mn@doi [\mnras] {10.1093/mnras/282.2.347}, \href {http://ads.nao.ac.jp/abs/1996MNRAS.282..347M} {282, 347}

\bibitem[\protect\citeauthoryear{{Mulchaey}}{{Mulchaey}}{2000}]{Mulchaey2000}
{Mulchaey} J.~S.,  2000, \mn@doi [\araa] {10.1146/annurev.astro.38.1.289}, \href {https://ui.adsabs.harvard.edu/abs/2000ARA&A..38..289M} {38, 289}

\bibitem[\protect\citeauthoryear{{Munari}, {Biviano}, {Borgani}, {Murante}  \& {Fabjan}}{{Munari} et~al.}{2013}]{Munari2013}
{Munari} E.,  {Biviano} A.,  {Borgani} S.,  {Murante} G.,   {Fabjan} D.,  2013, \mn@doi [\mnras] {10.1093/mnras/stt049}, \href {https://ui.adsabs.harvard.edu/abs/2013MNRAS.430.2638M} {430, 2638}

\bibitem[\protect\citeauthoryear{{Musso} \& {Sheth}}{{Musso} \& {Sheth}}{2012}]{Musso2012}
{Musso} M.,  {Sheth} R.~K.,  2012, \mn@doi [\mnras] {10.1111/j.1745-3933.2012.01266.x}, \href {https://ui.adsabs.harvard.edu/abs/2012MNRAS.423L.102M} {423, L102}

\bibitem[\protect\citeauthoryear{{Musso}, {Cadiou}, {Pichon}, {Codis}, {Kraljic}  \& {Dubois}}{{Musso} et~al.}{2018}]{Musso2018}
{Musso} M.,  {Cadiou} C.,  {Pichon} C.,  {Codis} S.,  {Kraljic} K.,   {Dubois} Y.,  2018, \mn@doi [\mnras] {10.1093/mnras/sty191}, \href {https://ui.adsabs.harvard.edu/abs/2018MNRAS.476.4877M} {476, 4877}

\bibitem[\protect\citeauthoryear{{Navarro}, {Frenk}  \& {White}}{{Navarro} et~al.}{1997}]{NFW1997}
{Navarro} J.~F.,  {Frenk} C.~S.,   {White} S. D.~M.,  1997, \mn@doi [\apj] {10.1086/304888}, \href {https://ui.adsabs.harvard.edu/abs/1997ApJ...490..493N} {490, 493}

\bibitem[\protect\citeauthoryear{{Osipkov}}{{Osipkov}}{1979}]{Osipkov79}
{Osipkov} L.~P.,  1979, Soviet Astronomy Letters, 5, 42

\bibitem[\protect\citeauthoryear{{Osmond} \& {Ponman}}{{Osmond} \& {Ponman}}{2004}]{Osmond2004}
{Osmond} J. P.~F.,  {Ponman} T.~J.,  2004, \mn@doi [\mnras] {10.1111/j.1365-2966.2004.07742.x}, \href {https://ui.adsabs.harvard.edu/abs/2004MNRAS.350.1511O} {350, 1511}

\bibitem[\protect\citeauthoryear{{Pacaud} et~al.,}{{Pacaud} et~al.}{2018}]{Pacaud2018}
{Pacaud} F.,  et~al., 2018, \mn@doi [\aap] {10.1051/0004-6361/201834022}, \href {https://ui.adsabs.harvard.edu/abs/2018A&A...620A..10P} {620, A10}

\bibitem[\protect\citeauthoryear{{Paranjape} \& {Padmanabhan}}{{Paranjape} \& {Padmanabhan}}{2017}]{Paranjape2017}
{Paranjape} A.,  {Padmanabhan} N.,  2017, \mn@doi [\mnras] {10.1093/mnras/stx659}, \href {https://ui.adsabs.harvard.edu/abs/2017MNRAS.468.2984P} {468, 2984}

\bibitem[\protect\citeauthoryear{{Paranjape}, {Hahn}  \& {Sheth}}{{Paranjape} et~al.}{2018}]{Paranjape2018}
{Paranjape} A.,  {Hahn} O.,   {Sheth} R.~K.,  2018, \mn@doi [\mnras] {10.1093/mnras/sty496}, \href {https://ui.adsabs.harvard.edu/abs/2018MNRAS.476.3631P} {476, 3631}

\bibitem[\protect\citeauthoryear{{Pasini} et~al.,}{{Pasini} et~al.}{2022}]{Pasini2022}
{Pasini} T.,  et~al., 2022, \mn@doi [\aap] {10.1051/0004-6361/202141211}, \href {https://ui.adsabs.harvard.edu/abs/2022A&A...661A..13P} {661, A13}

\bibitem[\protect\citeauthoryear{{Peirani} et~al.,}{{Peirani} et~al.}{2019}]{Peirani2019}
{Peirani} S.,  et~al., 2019, \mn@doi [\mnras] {10.1093/mnras/sty3475}, \href {https://ui.adsabs.harvard.edu/abs/2019MNRAS.483.4615P} {483, 4615}

\bibitem[\protect\citeauthoryear{{Pillepich}, {Porciani}  \& {Reiprich}}{{Pillepich} et~al.}{2012}]{Pillepich2012}
{Pillepich} A.,  {Porciani} C.,   {Reiprich} T.~H.,  2012, \mn@doi [\mnras] {10.1111/j.1365-2966.2012.20443.x}, \href {https://ui.adsabs.harvard.edu/abs/2012MNRAS.422...44P} {422, 44}

\bibitem[\protect\citeauthoryear{{Pizzuti}, {Saltas}, {Biviano}, {Mamon}  \& {Amendola}}{{Pizzuti} et~al.}{2022}]{Pizzuti2022}
{Pizzuti} L.,  {Saltas} I.~D.,  {Biviano} A.,  {Mamon} G.,   {Amendola} L.,  2022, arXiv e-prints, \href {https://ui.adsabs.harvard.edu/abs/2022arXiv220107194P} {p. arXiv:2201.07194}

\bibitem[\protect\citeauthoryear{{Planck Collaboration} et~al.,}{{Planck Collaboration} et~al.}{2016}]{Planck2016}
{Planck Collaboration} et~al., 2016, \mn@doi [\aap] {10.1051/0004-6361/201525830}, \href {https://ui.adsabs.harvard.edu/abs/2016A&A...594A..13P} {594, A13}

\bibitem[\protect\citeauthoryear{{Ponman}, {Bourner}  \& {Ebeling}}{{Ponman} et~al.}{1996}]{Ponman1996}
{Ponman} T.~J.,  {Bourner} P.~D.~J.,   {Ebeling} H.,  1996, in {Zimmermann} H.~U.,  {Tr{\"u}mper} J.,   {Yorke} H.,  eds, Roentgenstrahlung from the Universe. pp 357--360

\bibitem[\protect\citeauthoryear{{Popesso}, {Biviano}, {B{\"o}hringer}, {Romaniello}  \& {Voges}}{{Popesso} et~al.}{2005}]{Popesso2005}
{Popesso} P.,  {Biviano} A.,  {B{\"o}hringer} H.,  {Romaniello} M.,   {Voges} W.,  2005, \mn@doi [\aap] {10.1051/0004-6361:20041915}, \href {https://ui.adsabs.harvard.edu/abs/2005A&A...433..431P} {433, 431}

\bibitem[\protect\citeauthoryear{{Pratt}, {Croston}, {Arnaud}  \& {B{\"o}hringer}}{{Pratt} et~al.}{2009}]{pratt09}
{Pratt} G.~W.,  {Croston} J.~H.,  {Arnaud} M.,   {B{\"o}hringer} H.,  2009, \mn@doi [\aap] {10.1051/0004-6361/200810994}, \href {https://ui.adsabs.harvard.edu/abs/2009A&A...498..361P} {498, 361}

\bibitem[\protect\citeauthoryear{{Pratt}, {Arnaud}, {Biviano}, {Eckert}, {Ettori}, {Nagai}, {Okabe}  \& {Reiprich}}{{Pratt} et~al.}{2019}]{Pratt2019}
{Pratt} G.~W.,  {Arnaud} M.,  {Biviano} A.,  {Eckert} D.,  {Ettori} S.,  {Nagai} D.,  {Okabe} N.,   {Reiprich} T.~H.,  2019, \mn@doi [\ssr] {10.1007/s11214-019-0591-0}, \href {https://ui.adsabs.harvard.edu/abs/2019SSRv..215...25P} {215, 25}

\bibitem[\protect\citeauthoryear{{Predehl} et~al.,}{{Predehl} et~al.}{2021}]{Predehl2021}
{Predehl} P.,  et~al., 2021, \mn@doi [\aap] {10.1051/0004-6361/202039313}, \href {https://ui.adsabs.harvard.edu/abs/2021A&A...647A...1P} {647, A1}

\bibitem[\protect\citeauthoryear{{Ragagnin}, {Andreon}  \& {Puddu}}{{Ragagnin} et~al.}{2022}]{Ragagnin22}
{Ragagnin} A.,  {Andreon} S.,   {Puddu} E.,  2022, \mn@doi [\aap] {10.1051/0004-6361/202244397}, \href {https://ui.adsabs.harvard.edu/abs/2022A&A...666A..22R} {666, A22}

\bibitem[\protect\citeauthoryear{{Ramakrishnan}, {Paranjape}, {Hahn}  \& {Sheth}}{{Ramakrishnan} et~al.}{2019}]{Ramakrishnan2019}
{Ramakrishnan} S.,  {Paranjape} A.,  {Hahn} O.,   {Sheth} R.~K.,  2019, \mn@doi [\mnras] {10.1093/mnras/stz2344}, \href {https://ui.adsabs.harvard.edu/abs/2019MNRAS.489.2977R} {489, 2977}

\bibitem[\protect\citeauthoryear{{Robotham} et~al.,}{{Robotham} et~al.}{2011}]{Robotham2011}
{Robotham} A.~S.~G.,  et~al., 2011, \mn@doi [\mnras] {10.1111/j.1365-2966.2011.19217.x}, \href {https://ui.adsabs.harvard.edu/abs/2011MNRAS.416.2640R} {416, 2640}

\bibitem[\protect\citeauthoryear{{Robotham}, {Bellstedt}, {Lagos}, {Thorne}, {Davies}, {Driver}  \& {Bravo}}{{Robotham} et~al.}{2020}]{Robotham2020}
{Robotham} A.~S.~G.,  {Bellstedt} S.,  {Lagos} C. d.~P.,  {Thorne} J.~E.,  {Davies} L.~J.,  {Driver} S.~P.,   {Bravo} M.,  2020, \mn@doi [\mnras] {10.1093/mnras/staa1116}, \href {https://ui.adsabs.harvard.edu/abs/2020MNRAS.495..905R} {495, 905}

\bibitem[\protect\citeauthoryear{{Rozo} et~al.,}{{Rozo} et~al.}{2009}]{rozo2009}
{Rozo} E.,  et~al., 2009, \mn@doi [\apj] {10.1088/0004-637X/699/1/768}, \href {https://ui.adsabs.harvard.edu/abs/2009ApJ...699..768R} {699, 768}

\bibitem[\protect\citeauthoryear{{Rykoff} et~al.,}{{Rykoff} et~al.}{2008}]{rykoff2008}
{Rykoff} E.~S.,  et~al., 2008, \mn@doi [\apj] {10.1086/527537}, \href {https://ui.adsabs.harvard.edu/abs/2008ApJ...675.1106R} {675, 1106}

\bibitem[\protect\citeauthoryear{{Sabater} et~al.,}{{Sabater} et~al.}{2019}]{Sabater2019}
{Sabater} J.,  et~al., 2019, \mn@doi [\aap] {10.1051/0004-6361/201833883}, \href {https://ui.adsabs.harvard.edu/abs/2019A&A...622A..17S} {622, A17}

\bibitem[\protect\citeauthoryear{{Salvato} et~al.,}{{Salvato} et~al.}{2018}]{Salvato2018}
{Salvato} M.,  et~al., 2018, \mn@doi [\mnras] {10.1093/mnras/stx2651}, \href {https://ui.adsabs.harvard.edu/abs/2018MNRAS.473.4937S} {473, 4937}

\bibitem[\protect\citeauthoryear{{Salvato} et~al.,}{{Salvato} et~al.}{2022}]{Salvato2022}
{Salvato} M.,  et~al., 2022, \mn@doi [\aap] {10.1051/0004-6361/202141631}, \href {https://ui.adsabs.harvard.edu/abs/2022A&A...661A...3S} {661, A3}

\bibitem[\protect\citeauthoryear{{Schellenberger} \& {Reiprich}}{{Schellenberger} \& {Reiprich}}{2017}]{Schellenberger2017}
{Schellenberger} G.,  {Reiprich} T.~H.,  2017, \mn@doi [\mnras] {10.1093/mnras/stx1583}, \href {https://ui.adsabs.harvard.edu/abs/2017MNRAS.471.1370S} {471, 1370}

\bibitem[\protect\citeauthoryear{{Seppi} et~al.,}{{Seppi} et~al.}{2022}]{Seppi2022}
{Seppi} R.,  et~al., 2022, \mn@doi [\aap] {10.1051/0004-6361/202243824}, \href {https://ui.adsabs.harvard.edu/abs/2022A&A...665A..78S} {665, A78}

\bibitem[\protect\citeauthoryear{{Sheth}, {Mo}  \& {Tormen}}{{Sheth} et~al.}{2001}]{Sheth2001}
{Sheth} R.~K.,  {Mo} H.~J.,   {Tormen} G.,  2001, \mn@doi [\mnras] {10.1046/j.1365-8711.2001.04006.x}, \href {https://ui.adsabs.harvard.edu/abs/2001MNRAS.323....1S} {323, 1}

\bibitem[\protect\citeauthoryear{{Shi}, {Wang}  \& {Mo}}{{Shi} et~al.}{2015}]{Shi2015}
{Shi} J.,  {Wang} H.,   {Mo} H.~J.,  2015, \mn@doi [\apj] {10.1088/0004-637X/807/1/37}, \href {https://ui.adsabs.harvard.edu/abs/2015ApJ...807...37S} {807, 37}

\bibitem[\protect\citeauthoryear{{Smith}, {Cole}, {Baugh}, {Zheng}, {Angulo}, {Norberg}  \& {Zehavi}}{{Smith} et~al.}{2017}]{Smith2017}
{Smith} A.,  {Cole} S.,  {Baugh} C.,  {Zheng} Z.,  {Angulo} R.,  {Norberg} P.,   {Zehavi} I.,  2017, \mn@doi [\mnras] {10.1093/mnras/stx1432}, \href {https://ui.adsabs.harvard.edu/abs/2017MNRAS.470.4646S} {470, 4646}

\bibitem[\protect\citeauthoryear{{Strateva} et~al.,}{{Strateva} et~al.}{2001}]{Strateva2001}
{Strateva} I.,  et~al., 2001, \mn@doi [\aj] {10.1086/323301}, \href {https://ui.adsabs.harvard.edu/abs/2001AJ....122.1861S} {122, 1861}

\bibitem[\protect\citeauthoryear{{Sun}, {Voit}, {Donahue}, {Jones}, {Forman}  \& {Vikhlinin}}{{Sun} et~al.}{2009}]{Sun2009}
{Sun} M.,  {Voit} G.~M.,  {Donahue} M.,  {Jones} C.,  {Forman} W.,   {Vikhlinin} A.,  2009, \mn@doi [\apj] {10.1088/0004-637X/693/2/1142}, \href {https://ui.adsabs.harvard.edu/abs/2009ApJ...693.1142S} {693, 1142}

\bibitem[\protect\citeauthoryear{{Taylor} et~al.,}{{Taylor} et~al.}{2011}]{Tailor2011}
{Taylor} E.~N.,  et~al., 2011, \mn@doi [\mnras] {10.1111/j.1365-2966.2011.19536.x}, \href {https://ui.adsabs.harvard.edu/abs/2011MNRAS.418.1587T} {418, 1587}

\bibitem[\protect\citeauthoryear{{Tempel}, {Tuvikene}, {Kipper}  \& {Libeskind}}{{Tempel} et~al.}{2017}]{Tempel2017}
{Tempel} E.,  {Tuvikene} T.,  {Kipper} R.,   {Libeskind} N.~I.,  2017, \mn@doi [\aap] {10.1051/0004-6361/201730499}, \href {https://ui.adsabs.harvard.edu/abs/2017A&A...602A.100T} {602, A100}

\bibitem[\protect\citeauthoryear{{Tiret}, {Combes}, {Angus}, {Famaey}  \& {Zhao}}{{Tiret} et~al.}{2007}]{Tiret+07}
{Tiret} O.,  {Combes} F.,  {Angus} G.~W.,  {Famaey} B.,   {Zhao} H.~S.,  2007, \mn@doi [\aap] {10.1051/0004-6361:20078569}, 476, L1

\bibitem[\protect\citeauthoryear{{Vulic} et~al.,}{{Vulic} et~al.}{2022}]{Vulic2022}
{Vulic} N.,  et~al., 2022, \mn@doi [\aap] {10.1051/0004-6361/202141641}, \href {https://ui.adsabs.harvard.edu/abs/2022A&A...661A..16V} {661, A16}

\bibitem[\protect\citeauthoryear{{Wang} et~al.,}{{Wang} et~al.}{2014}]{Wang2014}
{Wang} L.,  et~al., 2014, \mn@doi [\mnras] {10.1093/mnras/stt2481}, \href {https://ui.adsabs.harvard.edu/abs/2014MNRAS.439..611W} {439, 611}

\bibitem[\protect\citeauthoryear{{Wechsler}, {Bullock}, {Primack}, {Kravtsov}  \& {Dekel}}{{Wechsler} et~al.}{2002}]{Wechsler2002}
{Wechsler} R.~H.,  {Bullock} J.~S.,  {Primack} J.~R.,  {Kravtsov} A.~V.,   {Dekel} A.,  2002, \mn@doi [\apj] {10.1086/338765}, \href {https://ui.adsabs.harvard.edu/abs/2002ApJ...568...52W} {568, 52}

\bibitem[\protect\citeauthoryear{{Wechsler}, {Zentner}, {Bullock}, {Kravtsov}  \& {Allgood}}{{Wechsler} et~al.}{2006}]{wechsler06}
{Wechsler} R.~H.,  {Zentner} A.~R.,  {Bullock} J.~S.,  {Kravtsov} A.~V.,   {Allgood} B.,  2006, \mn@doi [\apj] {10.1086/507120}, \href {http://adsabs.harvard.edu/abs/2006ApJ...652...71W} {652, 71}

\bibitem[\protect\citeauthoryear{{Weinmann}, {van den Bosch}, {Yang}  \& {Mo}}{{Weinmann} et~al.}{2006}]{Weinmann2006}
{Weinmann} S.~M.,  {van den Bosch} F.~C.,  {Yang} X.,   {Mo} H.~J.,  2006, \mn@doi [\mnras] {10.1111/j.1365-2966.2005.09865.x}, \href {https://ui.adsabs.harvard.edu/abs/2006MNRAS.366....2W} {366, 2}

\bibitem[\protect\citeauthoryear{{Williams} et~al.,}{{Williams} et~al.}{2019}]{Williams2019}
{Williams} W.~L.,  et~al., 2019, \mn@doi [\aap] {10.1051/0004-6361/201833564}, \href {https://ui.adsabs.harvard.edu/abs/2019A&A...622A...2W} {622, A2}

\bibitem[\protect\citeauthoryear{Yang, Mo, van~den Bosch, Pasquali, Li  \& Barden}{Yang et~al.}{2007}]{2007ApJ...671..153Y}
Yang X.,  Mo H.~J.,  van~den Bosch F.~C.,  Pasquali A.,  Li C.,   Barden M.,  2007, The Astrophysical Journal, 671, 153

\bibitem[\protect\citeauthoryear{{van der Marel}, {Magorrian}, {Carlberg}, {Yee}  \& {Ellingson}}{{van der Marel} et~al.}{2000}]{vanderMarel+00}
{van der Marel} R.~P.,  {Magorrian} J.,  {Carlberg} R.~G.,  {Yee} H.~K.~C.,   {Ellingson} E.,  2000, \aj, 119, 2038

\makeatother
\end{thebibliography}




\appendix


\bsp	
\label{lastpage}
\end{document}